\documentclass[12pt]{article}
\usepackage{amsmath}
\usepackage{graphicx,psfrag,epsf}
\usepackage{enumerate}
\usepackage{natbib}
\usepackage{amsthm,color}
\usepackage{amssymb}
\usepackage{latexsym}
\usepackage{float}
\usepackage{amsfonts}
\usepackage{bbm}
\usepackage{longtable}
\usepackage{booktabs}
\usepackage{multirow}

\usepackage[footnotesize]{caption}
\usepackage{wrapfig}
\usepackage{setspace,verbatim}

\newcommand{\blind}{1}

\renewcommand{\arraystretch}{.65}

\def\bP{\mbox{\boldmath $P$}}
\def\bQ{\mbox{\boldmath $Q$}}
\def\bD{\mbox{\boldmath $D$}}
\def\bsQ{\mbox{\boldmath \scriptsize $Q$}}
\def\th{^{\mbox{\scriptsize th}}}
\def\by{\mbox{\boldmath $y$}}
\def\b\pi{\mbox{\boldmath $\pi$}}

\def\pib{\mbox{\scriptsize pib}}
\def\thick{\mbox{\scriptsize thick}}
\def\mmse{\mbox{\scriptsize mmse}}

\begin{document}

\pagestyle{empty}


\begin{center}
{\singlespacing
\begin{Large}{\bf
A Bayesian Approach to Multi-State Hidden Markov Models: Application to Dementia Progression\\}
\vspace{.25in}
\end{Large}

\if1\blind
{
Jonathan P Williams$^{\dag\:\ddag}$, Curtis B Storlie$^\dag$, Terry M Therneau$^\dag$, Clifford R Jack Jr$^\dag$, \\ Jan Hannig$^\ddag$\\[.15in]
$^\dag$Mayo Clinic\\
$^\ddag$University of North Carolina at Chapel Hill\\
} \fi
 
\begin{abstract}
People are living longer than ever before, and with this arises new complications and challenges for humanity.  Among the most pressing of these challenges is of understanding the role of aging in the development of dementia.  This paper is motivated by the Mayo Clinic Study of Aging data for 4742 subjects since 2004, and how it can be used to draw inference on the role of aging in the development of dementia.  We construct a hidden Markov model (HMM) to represent progression of dementia from states associated with the buildup of amyloid plaque in the brain, and the loss of cortical thickness.  A hierarchical Bayesian approach is taken to estimate the parameters of the HMM with a truly time-inhomogeneous infinitesimal generator matrix, and response functions of the continuous-valued biomarker measurements are cut-point agnostic.  A Bayesian approach with these features could be useful in many disease progression models.  Additionally, an approach is illustrated for correcting a common bias in delayed enrollment studies, in which some or all subjects are not observed at baseline.  Standard software is incapable of accounting for this critical feature, so code to perform the estimation of the model described below is made available online.

\vspace{.06in}
\noindent
{\em Keywords}: Hierarchical Bayesian Modeling; Population Study; Hidden Markov Model; Alzheimer's Disease; Death Bias.

\vspace{.06in}
\noindent
{\em Running title}: Multi-State Hidden Markov Model for Dementia Progression

\if1\blind
{
\vspace{.06in}
\noindent
{\em Corresponding Author}: Jonathan Williams, \verb5williams.jonathan1@mayo.edu5
} \fi
\end{abstract}
}
\end{center}

\newpage

\pagestyle{plain}
\setcounter{page}{1}

\vspace{-.25in}
\section{Introduction}\label{introMayo}
\vspace{-.09in}

People are living longer and with this arises new complications and challenges for humanity.  Among the most pressing of these challenges is of understanding the role of aging in the development of dementia.  Such is the initiative of the Mayo Clinic Study of Aging (MCSA), a large prospective study with the goal of understanding the natural history of dementia and particularly Alzheimer's Disease.  

This paper is motivated entirely by the MCSA and how the resulting data can be used to draw inference on the role of aging in the development of dementia.  The goal is to create a model of progression to dementia which can accommodate: (1) A wide variation in age (the dominant variable under consideration), (2) Significant fluctuation in the time between subject visits, (3) Different amount of information available for each subject (e.g., missing visits and/or clinical data), and (4) Subject specific covariates.  


The main contribution of this work is to provide an innovative statistical analysis of this important and unique data set via a continuous-time, discrete-state hidden Markov model (HMM) estimated within the Bayesian paradigm.  Additionally, we demonstrate the existence of and provide solutions for various methodological gaps in the analysis of disease progression for studies like the MCSA.  First, we provide an approach for correcting a common bias in {\it delayed enrollment} studies which has been overlooked in the literature.  Second, we introduce a methodological framework for estimating the strength and persistence of a separate {\it death rate bias} specific to death rates, which could be present in any study relying on enrollment of subjects.  Our final methodological innovation is a proposed Bayesian approach to estimating the biomarker regions most associated with high/low burden states in a manner that does not require the specification of cut-points.

The term {\it delayed enrollment}, here, is used to describe a study with a given baseline (age 50 in the case of the MCSA) such that some or all subjects are $not$ observed at baseline.  We demonstrate empirically that the effects of this bias cannot be ignored, and existing software is not equipped to handle this feature.

We formalize the discrete-state space exhibited in Figure \ref{StateSpace} in which many of the states are defined by continuous biomarkers.  The previous work of \cite{Jack2016} defined a state space similar to Figure \ref{StateSpace}, but in which the high/low burden biomarker states were defined by practitioner chosen, hard biomarker cut-points.  Hard cut-points for discretizing continuous measurements of biological processes are practically and philosophically problematic, and have to be chosen more or less arbitrarily.  

Moreover, we illustrate a general and effective framework for fitting a continuous-time, discrete-state HMM within the Bayesian paradigm, and the infinitesimal generator matrix of the underlying Markov process is allowed to be truly time-inhomogeneous (as a function of an individual's age).  Time must be treated as continuous because, as in much of medical research, subjects are often observed irregularly in time.  

Our final contribution is that in addition to the effect of age, the effects of the covariates gender, number of years of education, and presence of an APOE-$\varepsilon$4 allele on the infinitesimal transition rates are also estimated.  The importance of these variables has been well documented in the medical literature but their effect on aging has not been studied in this context (i.e., how they affect the transition rates between states in Figure~\ref{StateSpace}).  In addition to the new insights these features bring to the medical community, flexible software to fit the models described below is provided at \if1\blind{\verb1https://jonathanpw.github.io/software.html1}\fi \if0\blind{*blinded*}\fi.


Our analysis builds on the work of \cite{Jack2016} with more sophisticated modeling which allows for deeper insights.  They found that a Markov model of disease progression for dementia is indeed a natural approach, that almost all rates are log-linear, and at age 50 nearly everyone is in state A$-$N$-$ (i.e., low Amyloid burden and low cortical thickness loss burden which is state 1 in Figure \ref{StateSpace}) but that soon begins to change.

Most implementations of a continuous-time, discrete-state HMM, including \cite{Jack2016} estimate parameters in a maximum likelihood fashion.  However, as mentioned in \cite{Jack2016} optimization becomes exceedingly difficult as more parameters are introduced in the model.  Convergence time for standard methods may become impractical, and analytical gradient formulas for use in more efficient optimization procedures can become intractable.  Additionally, it is often difficult/awkward to fit prior information into an optimization-based frequentist approach (e.g., via constrained optimization or penalty functions which require tuning parameters), and deriving confidence intervals becomes a challenge.  Further, as the model becomes more complex to better capture reality in our application, prior information becomes necessary for practical identifiability of HMM parameters which makes Bayesian methodology a natural approach.  For these reasons, we propose a hierarchical Bayesian framework with model estimation via Markov chain Monte Carlo (MCMC).  Using MCMC for the estimation of complicated models requires creative proposal strategies, but is extremely flexible for a variety of model specifications.  Moreover, credible regions become a convenient way to represent uncertainty.  

Accounts of continuous-time, discrete-state HMM are given by \cite{Lange2013, Bureau2003, Jackson2003, Titman2008, Jack2016}, and within a Bayesian framework, \cite{Zhao2016}.  Further, \cite{Satten1996} is a very complete account of how to implement such a model, and is recommended reading for anyone not familiar with the methodology.  

More relevant to the present application, the work of \cite{Jackson2003} uses an HMM to model the state misclassification error of a disease, and includes age as a covariate on the transition rates.  However, they make a very restrictive assumption that the transition rates are constant between subject observation times.  The work of \cite{Jack2016} is seemingly the first in the literature to estimate the transition rates as a function of age, in a truly continuous-time fashion for a multi-state model of dementia, however, \cite{Yu2010} also treated transition probabilities as a function of age, in a discrete-time fashion.


The organization of the rest of the paper is as follows.  Section \ref{methodsMayo} describes the MCSA data set and the HMM methodology.  An illustration of the {\it death rate bias} and the {\it delayed enrollment bias} is then given in Section \ref{population_challenges}, and a simulation study is provided in Section \ref{demData} which implements our Bayesian estimation procedure on synthetic data generated to resemble the MCSA data set.  Finally, a detailed analysis of the MCSA data is presented in Section~\ref{RealDataMayo}, accompanied by a discussion and interpretation of the biological findings.  This paper also has accompanying supplementary material containing details of the computations and further MCSA analysis results.

\vspace{-.25in}
\section{Methodology}\label{methodsMayo}
\vspace{-.15in}

\subsection{Description of the Mayo Clinic Study of Aging data}
\vspace{-.07in}

The MCSA has enrolled a large age/sex stratified random sample from Olmsted County, Minnesota.  Subjects are followed forward approximately every 15 months, and clinical visits collect information on major aspects of the disease.  The MCSA study began in 2004, currently has 4742 subjects, and is still ongoing (at the time of this writing) and enrolling new subjects.  Estimates of quantities such as expected time in a given state, probability of ever entering a given state, or the fraction of the population that will pass through a high amyloid burden state on the way to dementia, are all of clinical interest.  

It is known in the medical community that amyloid protein buildup in the brain, and significant neuro-degeneration are strongly associated with dementia.  Accordingly, amyloid buildup, as measured by Pittsburgh Compound B (PIB) from a Positron Emission Tomography (PET) scan, and neuro-degeneration, as measured by cortical thickness (Thickness) from Magnetic Resonance Imaging (MRI), are continuous outcomes measured during the regular clinical visits for approximately 50 percent of the subjects.  

\begin{figure}[t]
\vspace{-.13in}
\centering
\includegraphics[scale=.48]{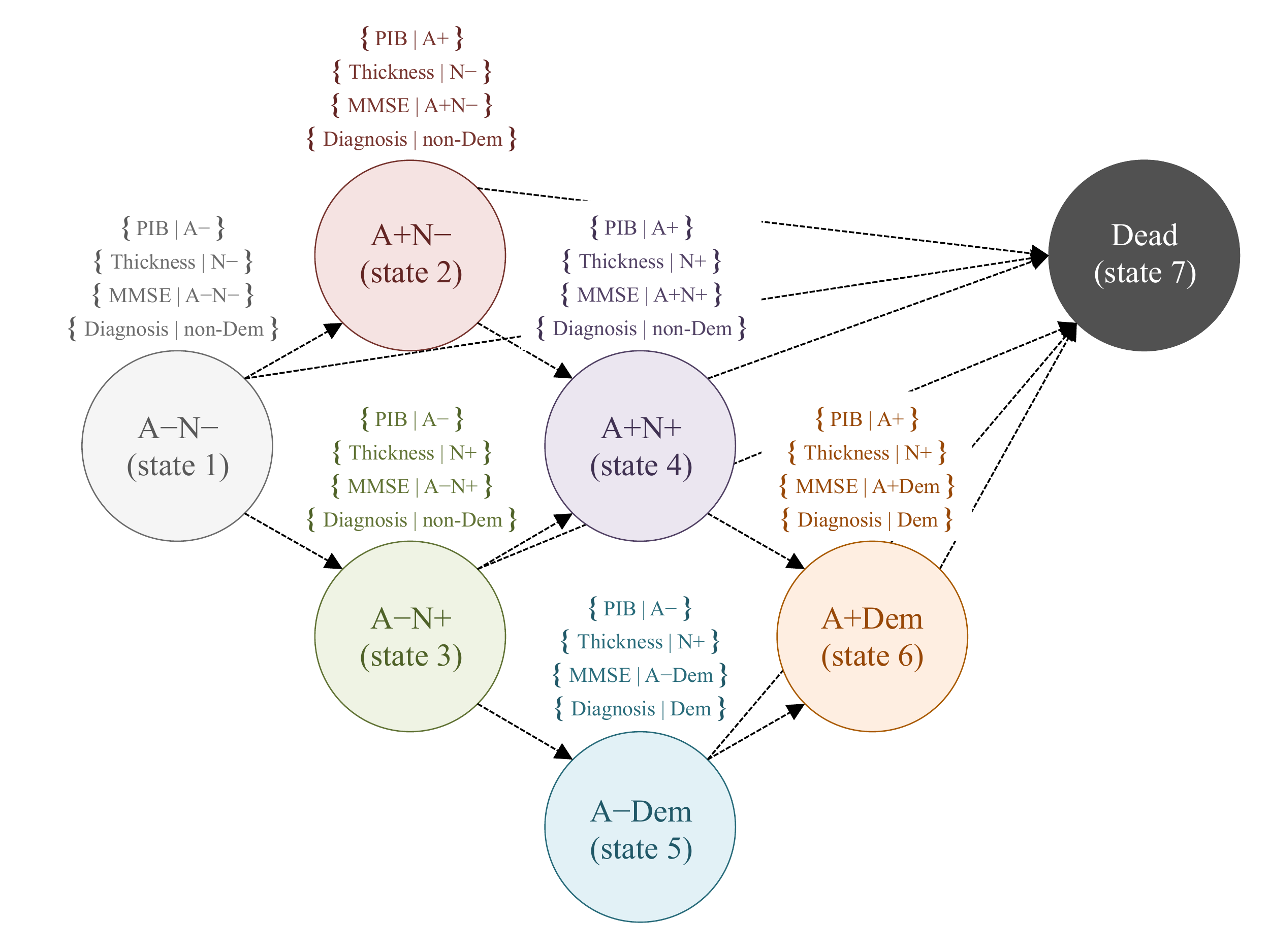}
\vspace{-.13in}
\caption{\footnotesize  $\!$State space.  Emitted response variables are displayed in brackets above the respective hidden state.  A+ corresponds to high amyloid burden, and N+ corresponds to high neuro-degenerative burden.  States 1-4 are all non-demented.}\label{StateSpace}
\end{figure}

With regard to dementia, high amyloid burden is a notion which refers to a build-up of amyloid plaques in the brain significant enough to effect pathways and lead to neuro-degeneration, but precise measurements of the extent of amyloid protein would require autopsy (PIB measurements serve as a proxy for measuring this extent).  Likewise, high neuro-degeneration burden refers to a state of loss of neurons and synapses denoted by atrophy of the cerebral cortex in Alzheimer's-sensitive areas. 

For the (approximately 50 percent of) subjects who were not chosen to undergo regular brain scans less clinical information is available.  However, the Mini Mental State Exam (MMSE) is almost always observed.  The MMSE is a questionnaire-based test administered by a medical professional to assess cognitive impairment on an integer scale out of 30 points \citep{Xu2015}.  Furthermore, baseline data such as age, sex, clinical and genetic markers is always recorded in the data.

Finally, at the time of observation all subjects are determined to be either cognitively unimpaired, or to be demented.  This represents a substantial amount of information for making inference on the underlying cognitive state of a subject.  However, diagnosing dementia is not an exact science, and so the observed label is not without error.

\vspace{-.15in}
\subsection{The HMM state space and emitted response variables}
\vspace{-.07in}

A simplistic formalization of the biology is to theorize a seven-state model to describe cognitive health in relation to dementia.  Figure \ref{StateSpace} illustrates such states, and depicts the allowed transitions with directional arrows.  A notable feature of the state space is that an individual must be in a high neuro-degeneration burden state (i.e., N+) to develop dementia, but not necessarily in a high amyloid burden state (i.e., A+).  In fact, the transition from A+N+ (state 4) to A+Dem (state 6) is identified as Alzheimer's Disease, a particular type of dementia.  Isolating this Alzheimer's Disease transition is not possible using the previous state space of \cite{Jack2016}.

Time must be treated as continuous because patient visit times are irregular, and the underlying sequence of states visited for an individual is hidden by uncertainty (even when PIB/Thickness are available, they are only proxies for the true level of amyloid/neuro-degeneration burden).  Moreover, the states of high amyloid burden and neuro-degenerative burden are not precisely defined and are best treated as hidden.  It is worth remarking that amyloid build-up and neuro-degeneration each develop on a continuum, but the time spent in any intermediate states, not explicitly represented by the state space in Figure \ref{StateSpace}, is believed to be relatively short and thus ignorable in these data.

The PIB and Thickness values associated with amyloid buildup and neuro-degeneration, respectively, are used as emitted response variables (in a traditional hidden Markov model sense) to make inference on an underlying sequence of states visited.  States 2, 4, and 6 which correspond to increased amyloid burden will emit PIB values from a distribution corresponding to A+, while states 1, 3, and 5 will emit PIB from the distribution corresponding to A$-$, and similarly for neuro-degenerative burden (N+ or N$-$).  

\begin{figure}[t]
\vspace{-.15in}
\centering
\includegraphics[scale=.39]{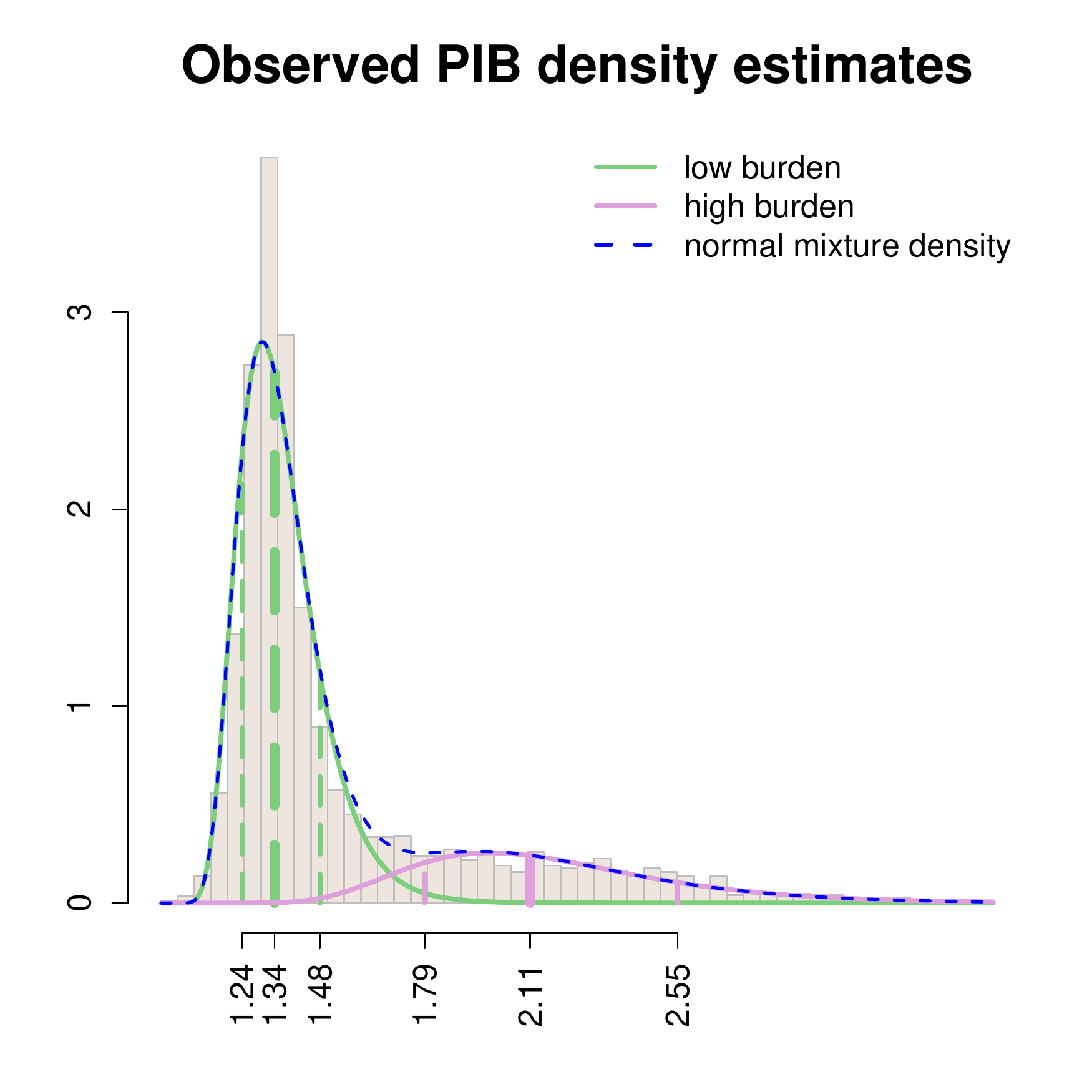}\includegraphics[scale=.39]{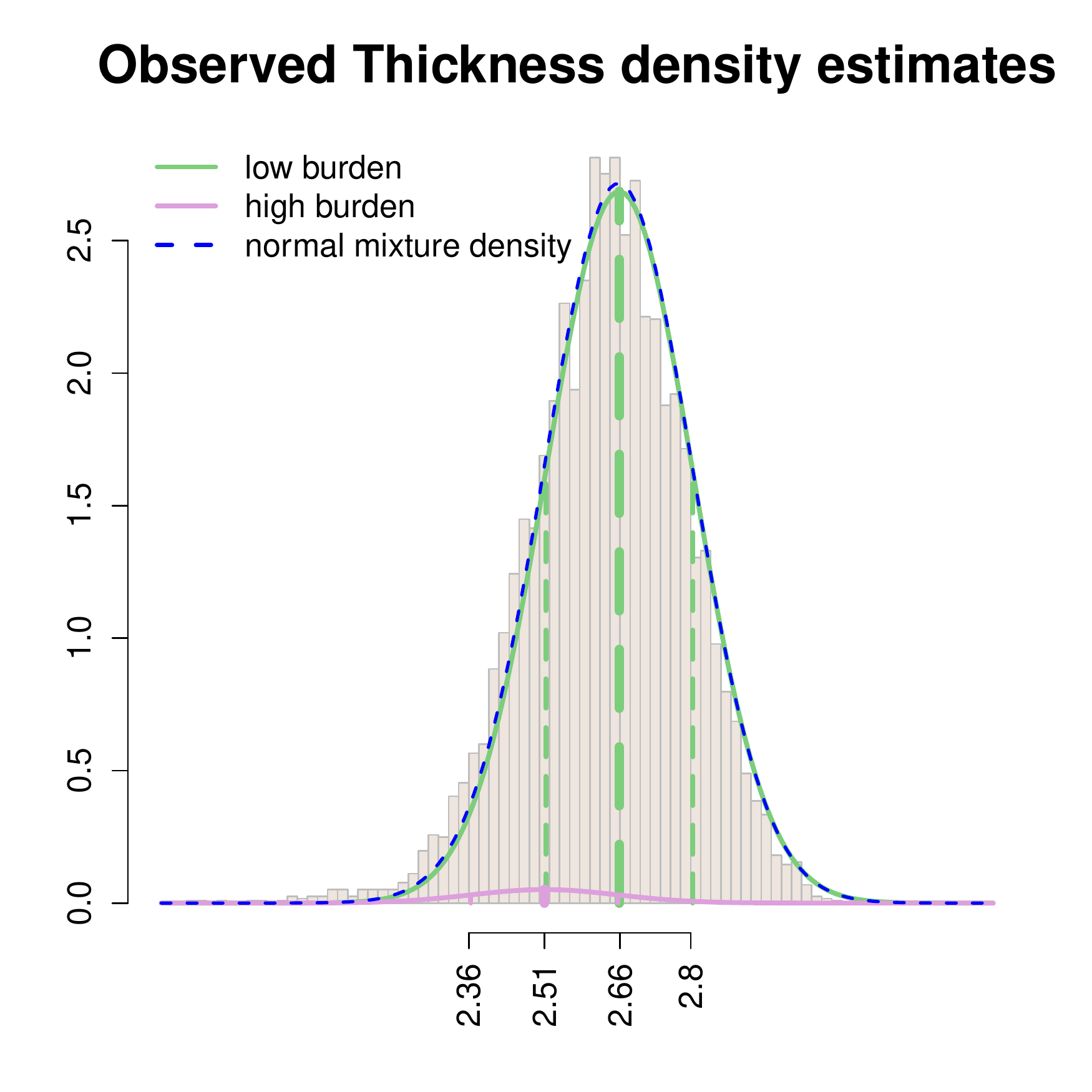}
\vspace{-.15in}
\caption{\footnotesize Observed response data for PIB which is a measure of amyloid buildup from a PET scan, and (cortical) Thickness which is associated with neuro-degeneration.  Note that the response densities for PIB correspond to the data transformation, $\log(\text{PIB}-1)$.  The component density estimates correspond to the posterior mean estimates from Section~\ref{RealDataMayo}.  The blue dashed lines represent the normal mixture density estimates.}\label{pibThicknessDensity}
\end{figure}

\begin{wrapfigure}{R}{.49\textwidth}
\vspace{-.15in}
\centering
\includegraphics[width=0.47\textwidth]{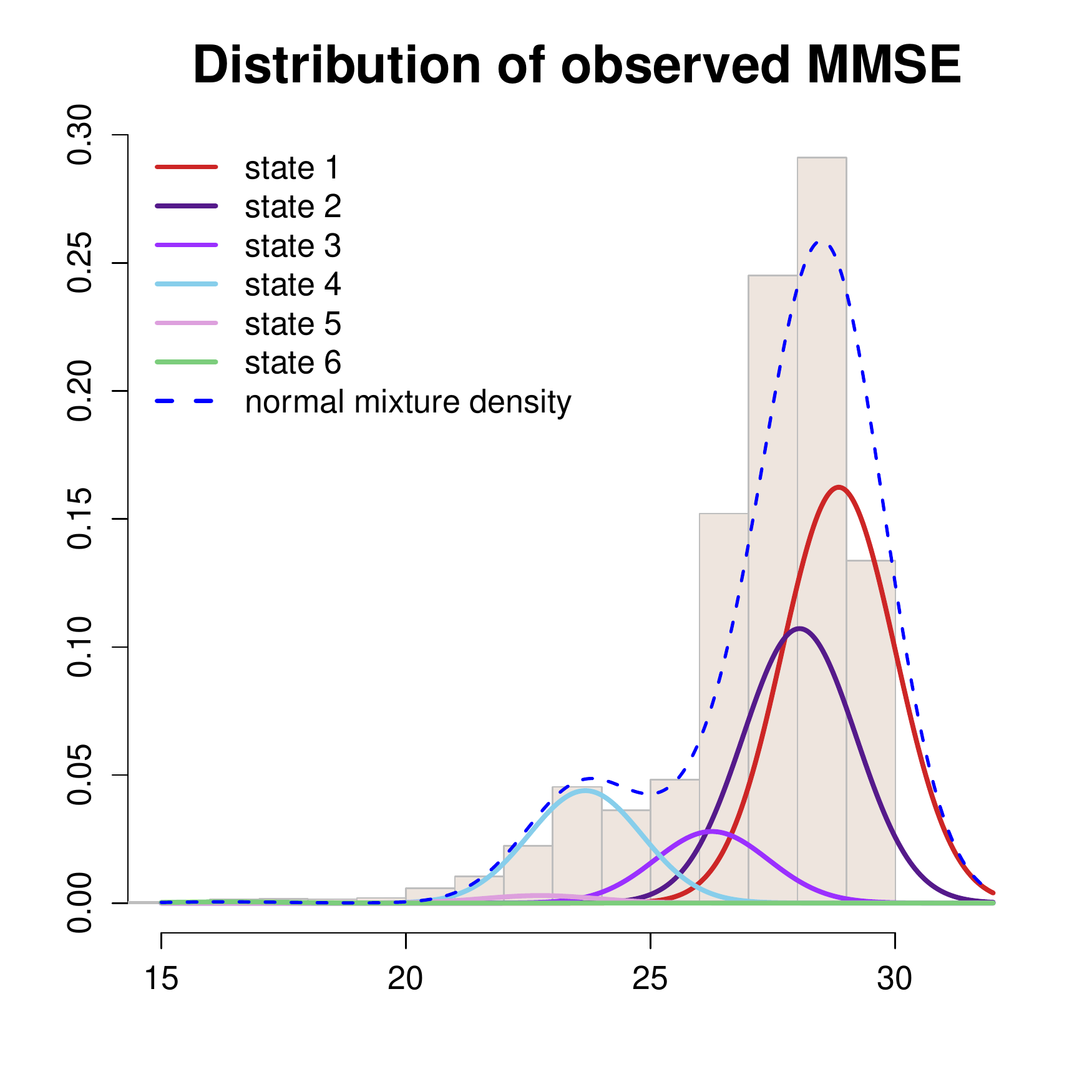}
\vspace{-.2in}
\caption{\footnotesize Observed response data for MMSE test scores associated with the six non-death states in the state space.  The component density estimates here correspond to the posterior mean estimates from Section~\ref{RealDataMayo}.  The blue dashed line represents the normal mixture density estimate.}\label{mmseDensity}
\end{wrapfigure}

The prior distributions for these response distribution parameters was chosen to correspond to biomarker values which are consistent with the medical community's most up-to-date understanding of the biology.  Using a Gaussian distribution for Thickness and for $\log(\text{PIB} - 1)$ appears to be quite reasonable, as there is evidence that the error from the PIB measurements follows more closely to a constant coefficient of variation than to a constant variance.   Figure~\ref{pibThicknessDensity} displays histograms of the observed response data along with the respective normal mixture densities resulting from the posterior mean estimates from the full HMM described in the coming sections. 

The MMSE score serves as an additional emitted response, and a separate Gaussian emission distribution for each of the first six states is assumed (deceased subjects do not emit cognitive test scores).  Figure \ref{mmseDensity} overlays the estimated six-component normal mixture density function on top of a histogram of the observed MMSE scores from the study subjects' visits.

Lastly, a simple misclassification response model is used to allow for a probability of a dementia diagnosis given the underlying state is a dementia state (i.e., states 5 or 6), and given it is not a dementia state.  Death is the only state in the state space which is known without error, and the exact time of death is known, as well.

\vspace{-.15in}
\subsection{Continuous-time transition probabilities}\label{ContTimeTrans}
\vspace{-.07in}

This section serves to specify the hidden Markov model in the context of the state space illustrated in Figure \ref{StateSpace}.  For $r, s \in \{1,2,3,4,5,6,7\}$ and $h, t \ge 0$, the probability of transitioning from state $r$ at time $h$ to state $s$ at time $h + t$ is denoted by $P_{r,s}(h,t) = P(S(h+t) = s | S(h) = r).$  Assuming these probabilities are differentiable functions in $t$ and that the Markov process is time-homogeneous, it can be shown that they satisfy the Kolmogorov forward equations (\cite{Karlin1981}),
\vspace{-.25in}\begin{equation}\label{forwardEq}
\bP'(t) = \bP(t)\bQ,
\vspace{-.25in}\end{equation}
where $\bQ$ is called the transition rate matrix, and $\bP(t)$ is the matrix with components
\vspace{-.25in}
\begin{equation}
P_{rs}(t) := P_{rs}(h=0,t).
\label{eq:prst}
\vspace{-.25in}
\end{equation}
Note that $h$ can be taken to be $0$ in (\ref{eq:prst}) because the probabilities are assumed for now to be time-homogeneous.  The off-diagonal components of $\bQ$ are interpreted as the change in transition probabilities for an infinitesimal amount of time into the future, i.e., 
\vspace{-.15in}\begin{equation}\label{InfinitesimalRate}
q_{rs} 
= \lim_{t\downarrow 0} \frac{P(S(t) = s | S(0) = r)}{t}, \ \ r \ne s, 
\vspace{-.15in}\end{equation}
with diagonal elements $q_{rr} = -\sum_{s\ne r}q_{rs}$. 

The forward equations in (\ref{forwardEq}) have the matrix exponential solution, $\bP(t) = e^{t \bsQ}$.  However, as discussed in Section \ref{introMayo}, the transition rates will be expressed as a function of a subject's age at the time of transition.  That is, $\bQ = \bQ(t)$ which violates the time-homogeneity of the Markov process.  A simple work around is to discretize the effect of age and assume that the transition rates only change when a subject's integer age changes.  Doing so implies that subjects' transition rates, $\bQ$, remain constant between birthdays and yields 
\vspace{-.15in}\begin{equation}\label{MatExp}
\bP(h,t) = e^{(\lfloor h+1\rfloor - h)\bsQ(\lfloor h\rfloor)}\cdot e^{\bsQ(\lfloor h + 1\rfloor)}\cdots e^{\bsQ(\lfloor h + t\rfloor - 1)}\cdot e^{(h+t-\lfloor h+t\rfloor)\bsQ(\lfloor h + t\rfloor)}, 
\vspace{-.15in}\end{equation}
for $\lfloor h\rfloor \ne \lfloor h + t\rfloor$, where $h$ represents the subject's current age, $t$ is the time (in years) into the future, and $\lfloor\cdot\rfloor$ is the floor function.

Observe that with expression (\ref{MatExp}) all transition probabilities can be computed, as long as the components of $\bQ(t)$ are specified.  As mentioned above, the transition rates will be modeled as a function of age, gender, presence of an APOE-$\varepsilon$4 allele, and number of years of education.  Specifically, denoting each of the 13 nonzero transition rates illustrated in Figure \ref{StateSpace} by $q_{l}$ for $l \in \{1,\dots,13\}$,
\vspace{-.15in}\begin{equation}\label{logRates}
\log(q_{l}) = \beta_{0}^{(l)} + \beta_{1}^{(l)}\cdot\text{age} + \beta_{2}^{(l)}\cdot\text{male} + \beta_{3}^{(l)}\cdot\text{educ} + \beta_{4}^{(l)}\cdot\text{apoe4},
\vspace{-.25in}\end{equation}
where,
\vspace{-.15in}\[\footnotesize
\bQ = 
\begin{pmatrix}
- q_{1} - q_{2} - q_{3} & q_{1} & q_{2} & 0 & 0 & 0 & q_{3} \\
0 & - q_{4} - q_{5} & 0 & q_{4} & 0 & 0 & q_{5} \\
0 & 0 & - q_{6} - q_{7} - q_{8} & q_{6} & q_{7} & 0 & q_{8} \\
0 & 0 & 0 & - q_{9} - q_{10} & 0 & q_{9} & q_{10} \\
0 & 0 & 0 & 0 & - q_{11} - q_{12} & q_{11} & q_{12} \\
0 & 0 & 0 & 0 & 0 & - q_{13} & q_{13} \\
0 & 0 & 0 & 0 & 0 & 0 & 0 \\
\end{pmatrix}.
\vspace{.1in}\]

\begin{wrapfigure}{R}{.45\textwidth}
\vspace{-0.2in}
\centering
\includegraphics[width=0.46\textwidth]{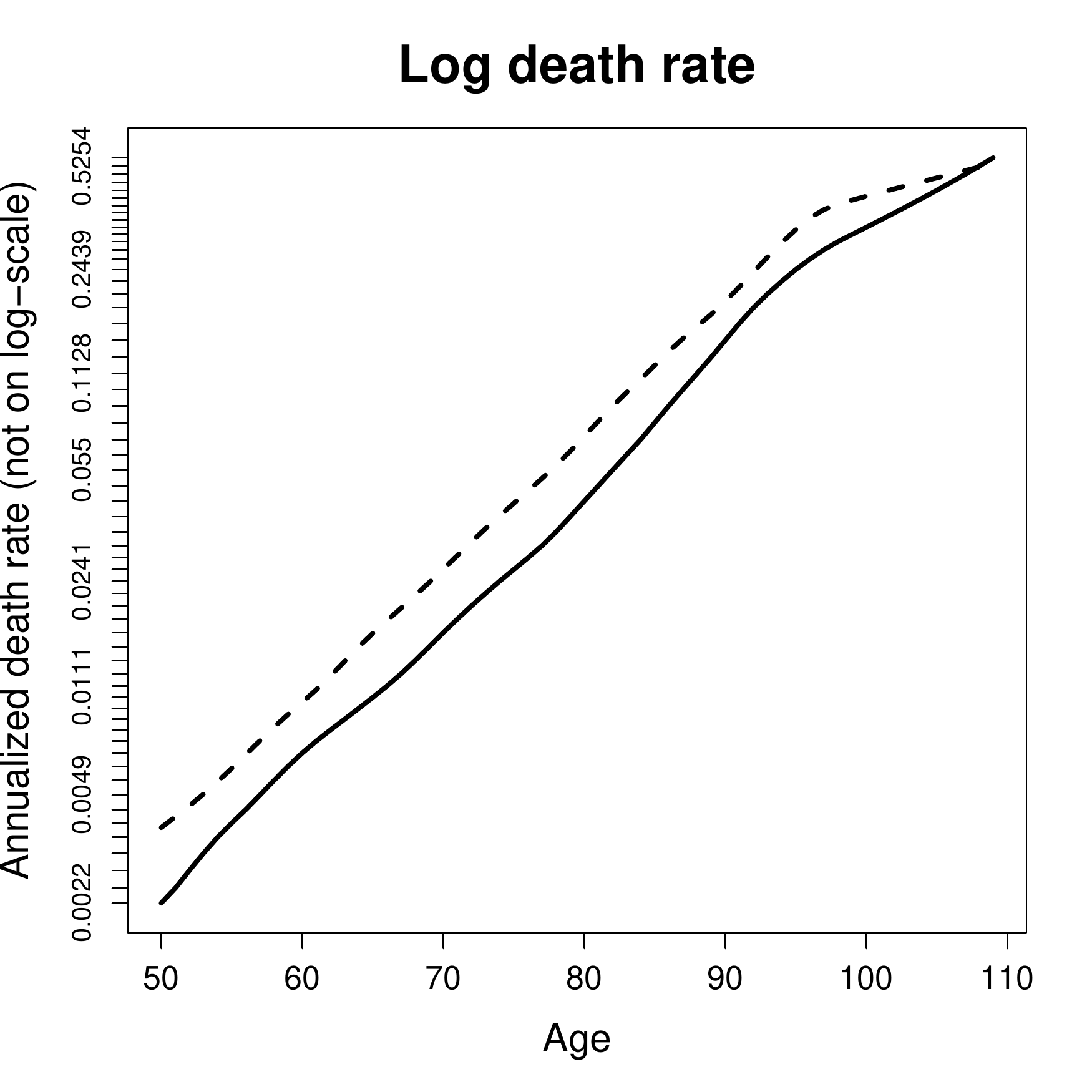}
\vspace{-.2in}
\caption{\footnotesize Annualized natural logarithm of Minnesota overall population death rates.  Solid line corresponds to female, and dashed line corresponds to male.}\label{logDeathRates}
\vspace{-.2in}
\end{wrapfigure}

\noindent
Overall death rates over age 50 are log-linear (see Figure \ref{logDeathRates}) which makes the log-linear function of age a natural starting place.  This functional form was also argued in \cite{Jack2016} to be reasonable for all of the rates except the rate from A$-$N$-$ (state 1) to A+N$-$ (state 2).  They compared log-linear rate to that obtained from log-cubic splines.  We came to the same conclusion and accordingly, a cubic spline is used for estimating only the rate of transition from state $1\rightarrow2$, with knots at ages 55, 65, 75, 90, and boundary knots at 50 and 120.

\vspace{-.15in}
\subsection{Likelihood function}\label{LikelihoodFun}
\vspace{-.07in}

If the states were known for each subject at each observation, then the contribution to the likelihood function from each subject consists of a product of matrix exponentials, i.e., apply (\ref{MatExp}) to get the probability of being in a given state for each observation time.  However, the underlying state sequences are not observed.  Within an HMM, rather, responses emitted from the underlying process (conditional on the true state of the process at a given point in time) are used to inform of the underlying state.  In this application, there are four emitted responses (i) $\log(\text{PIB} - 1)$, (ii) Thickness, (iii) MMSE (see Figures \ref{pibThicknessDensity} and \ref{mmseDensity}), and (iv) dementia diagnosis (binary).  Denote the observations for each of these four responses by $\by_{i,k} = [y_{i,k,1},y_{i,k,2},y_{i,k,3},y_{i,k,4}]'$, respectively, for the $i\th$ subject's $k\th$ clinical visit (observation).  Further, the $i\th$ subject has $n_{i}$ clinical visits, and has for each visit $k\in\{1,\dots,n_{i}\}$ an (unknown) state $s_{i,k}$.

Thus, the likelihood contribution from the $i\th$ subject at the $k\th$ visit can be expressed as
\vspace{-.15in}\[
f\big(\by_{i,k}\big)  =  \sum_{s_{i,k}=1}^7 f\big(\by_{i,k}, s_{i,k}\big)  =  \sum_{s_{i,k}=1}^7 P\big(s_{i,k}\big)\cdot f\big(\by_{i,k} \big| s_{i,k}\big),
\vspace{-.05in}\]
where the sum is taken over all possible states (since the true state sequence is unknown).  If a given response $y_{i,k,j}$ is missing (e.g., missing PIB scan), then the missing value is integrated out of the likelihood (i.e., the response density of the missing value contributes a 1 to the likelihood function).

Making the standard assumption that the responses are conditionally independent given an underlying state sequence $s_{i,1}, \dots, s_{i,n_{i}}$, and applying the Markov property for the state sequence gives,
\vspace{-.175in}\[
  f(\by_{i,1}, \dots, \by_{i,n_{i}}) =  \sum P(s_{i,1})P(s_{i,2} | s_{i,1}) \cdots P(s_{i,n_{i}} | s_{i,n_{i}-1}) \cdot \prod_{k=1}^{n_{i}} f(\by_{i,k} | s_{i,k}) \;\;\;\;\;\;\;\;\;\;\;\;\;\;\;\;\;\;\;\;\;\;\;\;\;\;\;\;\;\;\;\;\;\;\;\;
\]
\vspace{-.70in}\begin{eqnarray}
\;\;\;\;\;& = &\!\!\!\!\sum_{s_{i,1}=1}^{7} \!\!P(s_{i,1})f(\by_{i,1} | s_{i,1}) \cdot \!\sum_{s_{i,2}=1}^{7} \!\!P(s_{i,2} | s_{i,1})f(\by_{i,2} | s_{i,2})  \cdots \!\!\sum_{s_{i,n_{i}}=1}^{7}\!\!\!P(s_{i,n_{i}} | s_{i,n_{i}-1})f(\by_{i,n_{i}} | s_{i,n_{i}}) \nonumber \\
& = & \!\!\b\pi' \bD_{(i,1)}  \cdot \bP(t_{i,1}, t_{i,2}-t_{i,1})\bD_{(i,2)}\cdots \bP(t_{i,n_{i}-1}, t_{i,n_{i}}-t_{i,n_{i}-1})\bD_{(i,n_{i})} \cdot {\bf 1},   \label{eq:like} \\[-.45in] \nonumber
\end{eqnarray}
where $\b\pi$ is the initial state probability vector, $\bD_{(i,k)}$ is a diagonal matrix with diagonal components $f(\by_{i,k} | s_{i,k})$ for each $s_{i,k} \in \{1,\dots,7\}$ in the state space, $\bP(t_{i,k-1}, t_{i,k}-t_{i,k-1})$ is the transition probability matrix given in (\ref{MatExp}) with $t_{i,k}$ denoting the (continuous) age of subject $i$ at visit $k$, and ${\bf 1}$ is a column vector of ones.  There are three subtle, but important features of the MCSA data that need to be addressed which result in slight modifications of this likelihood function.

Finally, in order to complete the specification of the likelihood, the form of the response density $f(\by_{i,k})$ and the baseline state probability vector $\b\pi_0$ must be specified.  It is assumed that the four responses, $y_{i,k,1}=\log(\text{PIB} - 1)$, $y_{i,k,2}=\text{Thickness}$, $y_{i,k,3}=\text{MMSE}$, and $y_{i,k,4}=\text{Dementia}$, are conditionally independent given state $s_{i,k}$ and subject specific covariates.  As illustrated in Figure \ref{pibThicknessDensity}, the transformed PIB measurements are assumed to be generated according to two normal random variables with different means.  One of the means, say $\mu_{A-}$, corresponds to the distribution of transformed PIB measurements for an individual in a low amyloid burden state, and the other, say $\mu_{A+}$, corresponds to a high amyloid burden state.  Specifically,
\vspace{-.15in}\begin{equation}\label{pibEq}
f(y_{i,k,1} | s_{i,k}) = \text{N}(y_{i,k,1} | \mu_{A-}, \sigma_{\pib})\cdot I_{\left\{s_{i,k} \in \{1,3,5\}\right\}} + \text{N}(y_{i,k,1} | \mu_{A+}, \sigma_{\pib})\cdot I_{\left\{s_{i,k} \in \{2,4,6\}\right\}},
\vspace{-.15in}\end{equation}
where $I_A$ is the indicator function equal to 1 if $A$ and 0 otherwise.
The variance of both Gaussians are assumed to be equal to aid in identifiability of the two groups.  The density function for Thickness and MMSE are defined analogously.  

For the Thickness response variable,
\vspace{-.15in}\begin{equation}\label{thickEq}
f(y_{i,k,2} | s_{i,k}) = \text{N}(y_{i,k,2} | \mu_{N-}, \sigma_{\thick})\cdot I_{\left\{s_{i,k} \in \{1,2\}\right\}} + \text{N}(y_{i,k,2} | \mu_{N+}, \sigma_{\thick})\cdot I_{\left\{s_{i,k} \in \{3,4,5,6\}\right\}},
\vspace{-.15in}\end{equation}
and for the MMSE response variable,
\vspace{-.175in}\begin{equation}\label{mmseEq}
f(y_{i,k,3} | s_{i,k}) = \text{N}(y_{i,k,3} | \mu, \sigma_{\mmse}), 
\vspace{-.33in}\end{equation}
with 
\vspace{-.15in}\[
\mu = \sum_{j=1}^{6}\alpha_{j}\cdot I_{\{s_{i,k} = j\}} + \alpha_{7}\cdot \text{age} + \alpha_{8}\cdot \text{male} + \alpha_{9}\cdot \text{educ} + \alpha_{10}\cdot \text{apoe4} + \alpha_{11}\cdot \text{ntests}.
\vspace{-.15in}\]
The first four covariates are the same as those in Section \ref{ContTimeTrans}, and `ntests' is the number of times a subject has taken the MMSE by a given clinical visit.  It is observed in the medical practice that scores on the MMSE may improve as an individual becomes familiar with the exam, and so the `ntests' covariate is included to control for this effect.

The probability mass function for misdiagnosis of dementia is
\vspace{-.15in}\begin{equation}\label{demEq}
P(y_{i,k,4} | s_{i,k}) = \text{\small Bernoulli}(y_{i,k,4} | p_{0})\cdot I_{\left\{s_{i,k} \in \{1,2,3,4\}\right\}} + \text{\small Bernoulli}(1-y_{i,k,4} | p_{1})\cdot I_{\left\{s_{i,k} \in \{5,6\}\right\}},
\vspace{-.15in}\end{equation}
where $y_{i,k,4} = 1$ if the subject was diagnosed with dementia, and $y_{i,k,4} = 0$ if not.  Accordingly, $p_{0}$ and $p_{1}$ are misclassification probabilities, with $p_{0}$ the probability of an incorrect diagnosis of dementia, and $p_{1}$ the probability of an incorrect non-diagnosis of dementia.

Lastly, the baseline state probability vector is $\b\pi_{0} = [\pi_{0,1}, \pi_{0,2}, \pi_{0,3}, \pi_{0,4}, 0, 0, 0]'$ where $\pi_{0,j} = P(s_{i,1} = j)$ for $j \in \{1,2,3,4\}$, for all subjects, $i$.  As the MCSA did not enroll demented or deceased individuals, subjects would necessarily have been in states 1-4 at baseline.  Accordingly,  $\sum_{j=1}^{4}\pi_{0,j} = 1$, and the last three components of $\b\pi_{0}$ must be zero.

\vspace{-.1in}
\subsubsection{Treating time of death as known without error}
\vspace{-.05in}

One feature common to many population-based studies is that death is observed without error and time of death is known exactly, so the likelihood must be modified to account for this more precise information \citep{Satten1996}.  Suppose that subject $i$ transitions to death at time $t_{i,n_{i}}$.  The final term in the subject's contribution to the likelihood can be re-expressed as follows.  For a state $s_{i,n_{i}-1}$ at time $t_{i,n_{i}-1}$, and $\varepsilon \in (0, t_{i,n_{i}} - t_{i,n_{i}-1})$, let, 
\vspace{-.15in}\[
B(\varepsilon) := \{s_{i,n_{i}}(t_{i,n_{i}}) = 7, s(t_{i,n_{i}} - \varepsilon) < 7\}.
\vspace{-.15in}\]
Then $B := \bigcap_{\varepsilon}B(\varepsilon)$ is the event that the $i\th$ subject dies precisely at time $t_{i,n_{i}}$.  Further,
\vspace{-.15in}\[
P(B(\varepsilon) | s_{i,n_{i}-1}) = \sum_{s(t_{i,n_{i}}-\varepsilon)=1}^{6}P\big(s(t_{i,n_{i}}-\varepsilon) | s_{i,n_{i}-1}(t_{i,n_{i}-1})\big)\cdot P\big(s_{i,n_{i}}(t_{i,n_{i}}) = 7 | s(t_{i,n_{i}}-\varepsilon)\big).
\vspace{-.15in}\]
Thus, dividing both sides by $\varepsilon$ and taking the limit as $\varepsilon \to 0$ gives a likelihood function value of
\vspace{-.15in}\begin{equation}\label{TransToDeath}
\sum_{s(t_{i,n_{i}})=1}^{6}P(s(t_{i,n_{i}}) | s_{i,n_{i}-1})\cdot Q_{s,7}(\lfloor t_{i,n_{i}}\rfloor)
\vspace{-.01in}\end{equation}
evaluated at the event $\{B | s_{i,n_{i}-1}\}$.  The quantity in (\ref{TransToDeath}) is interpreted as the average probability of being in each of the first six states the instant prior to death, each weighted by the probability of transitioning to death at the next instant (given by the instantaneous transition rates $Q_{s,7}(\lfloor t_{i,n_{i}}\rfloor)$).  Note that the response functions are not needed/defined when $s_{i,n_{i}} = 7$ because death is assumed observed without error.

\vspace{-.25in}
\section{Population-based study challenges for an HMM}\label{population_challenges}
\vspace{-.10in}

The purpose of this section is to demonstrate the existence of and provide solutions for two critical methodological gaps in the analysis of disease progression for studies like the MCSA.  After describing these two overlooked sources of bias in the literature, Section \ref{cavData} serves to demonstrate empirically on a simple synthetic data set the impact that ensues when the {\it delayed enrollment bias} is not properly addressed.  We argue that the effects of this bias cannot be ignored, and existing software is not equipped to handle this feature.  

\vspace{-.15in}
\subsection{The {\it death rate bias}}\label{death_bias_section}
\vspace{-.07in}

A common feature of many human population based studies ({\it delayed enrollment} or not), is the following {\it death rate bias}.  In addition to not representing those members of the population who would have enrolled if they were not already dead (which is tied into the {\it delayed enrollment bias}), it often the case that people are much less likely to enroll in a study if they are very sick and/or dying.  As a result, the death rate for the sub-population of individuals who are most likely to enroll is probably smaller than the overall population death rate, for at least the first few years after enrollment into the study.  Consistent with the difference between this healthier sub-population and the overall population, this phenomenon will lead to a reduced estimate of the death rate and higher likelihoods of other paths in the state space.  This represents a bias with respect to the true parameter values of the overall population.  

Our proposed approach to correct for the {\it death rate bias} is to explicitly estimate the bias on the non-dementia to death rate.  This can be done in a linear fashion by including two additional parameters in the log-linear rate equation (\ref{logRates}).  The first, say $c \le 0$, will be for estimating the baseline effect of the {\it death rate bias}, and the second, say $d \ge 0$, will be for estimating the linear slope at which the {\it death rate bias} vanishes for every integer year in the study (since the time effect is discretized annually).  That is, equation (\ref{logRates}) for only the non-dementia to death rate ($l \in \{3,5,8,10\}$) becomes,
\vspace{-.15in}\begin{equation}\label{logRates_death}
\log(q_{l}) = \beta_{0}^{(l)} + \beta_{1}^{(l)}\cdot\text{age} + \beta_{2}^{(l)}\cdot\text{male} + \beta_{3}^{(l)}\cdot\text{educ} + \beta_{4}^{(l)}\cdot\text{apoe4} + g(\text{iyears}),
\vspace{-.15in}\end{equation}
where `iyears' is integer years enrolled, and $g(\text{iyears}) := \min\big\{ c + d\cdot\text{iyears}, 0\big\}$.  The {\it death rate bias} term is only allowed to decrease the non-dementia to death rate.  Further, the smallest root of $g$ is the duration for which the {\it death rate bias} persisted in the study.  

The coefficients in equation (\ref{logRates_death}) become identifiable due to strong prior information available for the overall population death rate.  We provide evidence that this bias exists in the MCSA data by appealing to the fact that $c$ and $d$ are both estimated to be nonzero (see Figure \ref{deathBiasSpline}), and by demonstrating in Section \ref{demData} that all parameters in equation (\ref{logRates_death}) can be accurately estimated in a similar, truth known synthetic data setting.

\vspace{-.15in}
\subsection{The {\it delayed enrollment bias}}\label{delayed_enrollment_section}
\vspace{-.07in}

The {\it delayed enrollment bias} occurs in situations in which the first observation time for the $i\th$ subject, $t_{i,1}$, is not necessarily equal to the baseline age of 50 years old (in the case of the MCSA).  In this case, the probability of transitioning from the (unknown) underlying state at baseline to the (unknown) underlying state at the first observation must be accounted for in the likelihood function.  That is, the initial state probability vector, $\b\pi$ in (\ref{eq:like}), is replaced with, 
\vspace{-.15in}\begin{equation}
\b\pi(t_{i,1}) = 
\left[ v_{1}(t_{i,1}) \:,\: v_{2}(t_{i,1}) \:,\: v_{3}(t_{i,1}) \:,\: v_{4}(t_{i,1}) \:,\: 0 \:,\: 0 \:,\: 0\right]'
\frac{1}{\sum_{j=1}^{4}v_{j}(t_{i,1})},
\label{eq:initial_prob}
\vspace{-.15in}\end{equation}
where $v(t_{i,1})' = \b\pi_{0}' P(50, t_{i,1}-50)$, and $\b\pi_0$ is the initial state probability vector for a subject at baseline.  The last three components are set equal to zero due to the fact that demented and dead subjects are not enrolled into the study.  

If the initial probabilities from baseline to enrollment age were not conditioned on the underlying states being non-demented and non-dead, then the transition rates (especially those to dementia and death) will exhibit strong downward bias (with respect to the true population parameter values) due to the fact that neither demented nor dead subjects are enrolled.  We refer to this bias as the {\it delayed enrollment bias}, and it will be illustrated empirically in Section \ref{cavData}.  Relevant standard software such as the \verb1msm1 package in $R$ \citep{Jackson2011} does allow for specification of a common baseline in a {\it delayed enrollment} study (after manually adding censored observations in the data set at baseline), but unfortunately does not offer the ability to perform conditioning (rescaling) on the initial state probability vector, as in (\ref{eq:initial_prob}).

\vspace{-.15in}
\subsection{Demonstrating the effect of a {\it delayed enrollment bias}}\label{cavData}
\vspace{-.07in}

The Cardiac Allograft Vasculopathy (CAV) data set from the \verb1msm1 package was collected from a study of the progression of CAV, which is a common cause of death after heart transplant \citep{Sharples2003}.  The state space as described in \cite{Jackson2011} includes four states labeled `no CAV' (state 1), `mild/moderate CAV' (state 2), `severe CAV' (state 3), and `death' (state 4), and forward-only transitions are assumed (patients can only get worse).  Observed remissions in the state of a patient is considered a result of misclassification error.  In the CAV data, baseline is defined as time of heart transplant.  All patients are observed at baseline, and in fact at baseline all patients are in state 1 because CAV does not develop immediately.  

In this simple data set, the only response variable is the observed state at each visit which is assumed to follow a categorical distribution.  In particular, given a patient is in state 1, 2, or 3 there is a nonzero probability of observing an adjacent state.  Additionally, as is the case for the MCSA, death and time of death are known without error.  Formally, given a true state (row) the response probability mass function takes the form given in Table \ref{RespCAV}, where the probabilities $p_{1}$, $p_{2}$, $p_{3}$, and $p_{4}$ are each interpreted as the probability a of particular misclassification.  Note that the rows must sum to one.

\begin{table}[h]
\footnotesize
\renewcommand{\arraystretch}{1.1} 
\begin{center}
\begin{tabular}{c c | c c c c}
\multicolumn{2}{r}{} & \multicolumn{4}{c}{Observed state} \\
& & no CAV & mild & severe & death \\
\cline{2-6}
\multirow{4}{*}{True state} & no CAV & $1 - p_{1}$ & $p_{1}$ & 0 & 0 \\
& mild & $p_{2}$ & $1 - p_{2} - p_{3}$ & $p_{3}$ & 0 \\
& severe & 0 & $p_{4}$ & $1 - p_{4}$ & 0 \\
& death & 0 & 0 & 0 & 1 
\end{tabular}
\caption{Misclassification response function.}\label{RespCAV}
\vspace{-.2in}
\end{center}
\renewcommand{\arraystretch}{.65} 
\end{table}

Time is treated as continuous and discretized annually.  Integer years since heart transplant, and gender are included as covariates on the transitions rates for the CAV data,
\vspace{-.15in}\begin{equation}\label{logRates_cav}
\log(q_{l}) = \beta_{0}^{(l)} + \beta_{1}^{(l)}\cdot\text{iyears} + \beta_{2}^{(l)}\cdot\text{sex},
\vspace{-.15in}\end{equation}
where,
\vspace{-.05in}\[\footnotesize
\bQ = 
\begin{pmatrix}
- q_{1} - q_{2} & q_{1} & 0 & q_{2} \\
0 & - q_{3} - q_{4} & q_{3} & q_{4} \\
0 & 0 & - q_{5} & q_{5} \\
0 & 0 & 0 & 0 \\
\end{pmatrix}.
\vspace{-.00in}\]

To replicate the CAV data in a `known truth' simulation, first the HMM parameter estimates were obtained using the \verb1msm1 package on the original CAV data.  As suggested in \cite{Jackson2011}, the `BFGS' quasi-Newton optimization algorithm was specified to estimate the HMM parameters with the `msm' function.  The parameter estimates from the CAV data set are then used as the `true' values in order to generate new data sets.

Important features of the MCSA as distinct from the CAV data are that not all subjects are in state 1 at baseline, and virtually none of the subjects are observed at baseline.  Thus, to more closely resemble the MCSA data, each patient in the simulated CAV data is generated with a baseline state according to the following distribution,
\begin{center}
\footnotesize
\begin{tabular}{r | c c c c}
$s_{i,1}$ & no CAV & mild & severe & death \\
\hline
$P(s_{i,1})$ & 0.95 & 0.04 & 0.01 & 0 \\
\end{tabular}$\;.$
\end{center}
See the Supplementary Material for complete detail on how synthetic data was generated.

\begin{figure}[t]
\centering
  \vspace{-.15in}
\includegraphics[trim=0cm 9cm 0cm 0cm, clip=true, scale=0.55,page=1]{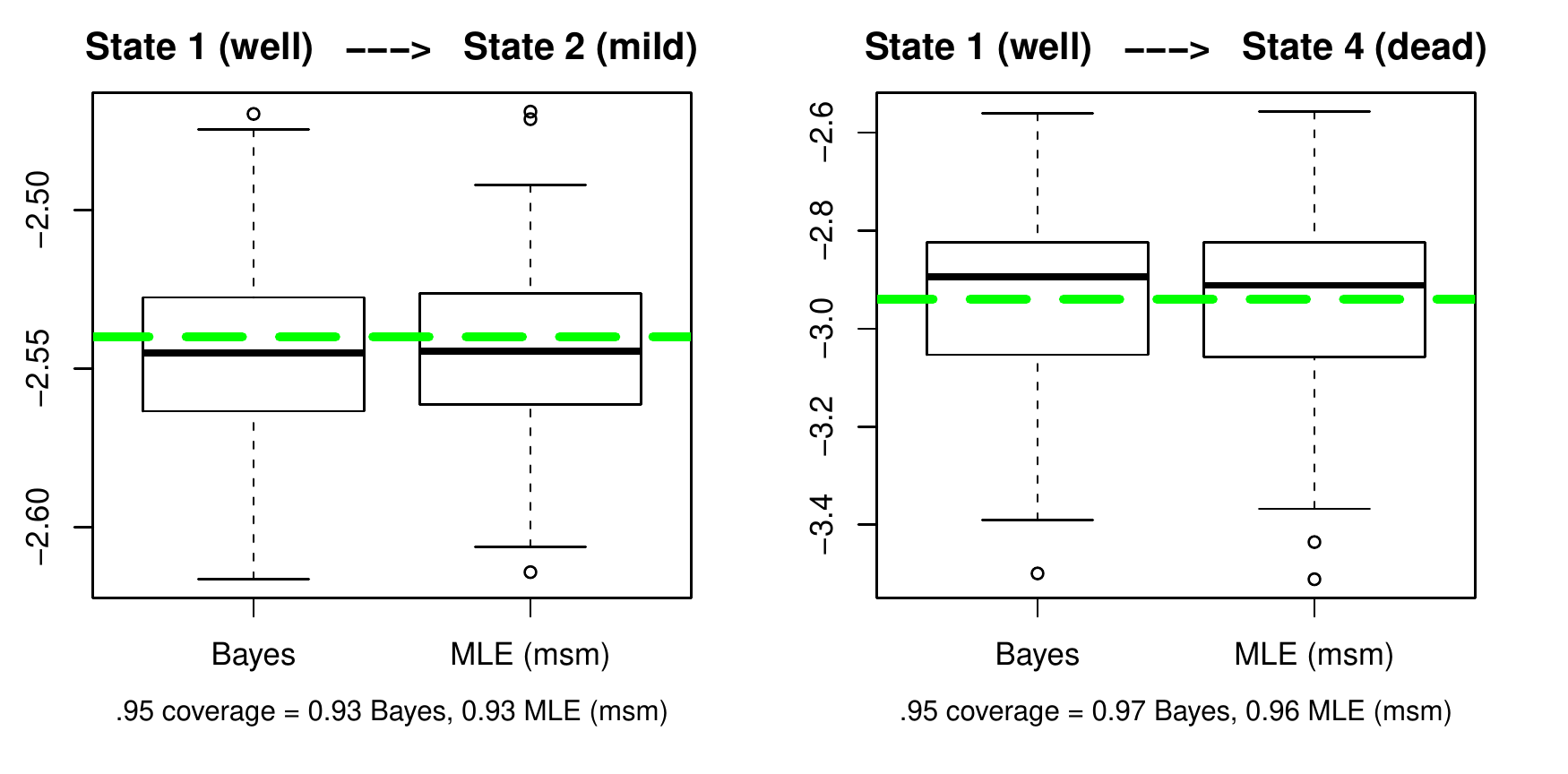}\includegraphics[trim=0cm 9cm 9cm 0cm, clip=true, scale=0.55,page=2]{qmat_intercept}
\includegraphics[trim=0cm 0cm 0cm 9cm, clip=true, scale=0.55,page=1]{qmat_intercept}
\vspace{-.2in}
\caption{Intercept coefficient estimates for a traditional study in which all subjects are enrolled at a common baseline time. `Bayes' corresponds to the Bayesian posterior mean estimates,  and `MLE (msm)' corresponds to maximum likelihood estimates computed from the `msm' function.  Green dashed lines represent the true values.  Coverage is the proportion of .95 probability credible intervals (confidence intervals for the MLE) which contain the true parameter value.}\label{cavSim}
\end{figure}

A sample size of 2000 was generated for 100 simulated data sets.  Figure~\ref{cavSim} shows estimates of the log-rate intercept coefficients from equation (\ref{logRates_cav}) using the posterior mean from the proposed Bayesian approach, and the MLE obtained via \verb1msm1.  The box plots are over the 100 estimates of each parameter and demonstrate that in a simple idealized setting the Bayesian and MLE estimates are very similar.

In the synthetic data example above, all patients are observed at baseline; this is the type of study for which the \verb1msm1 package was designed.  To illustrate the bias that ensues when subjects are not observed at baseline, the 100 data sets are generated once more from the same random generator seeds.  However, instead of beginning with the initial observations at baseline (zero  years after heart transplant), the time of initial observation is generated, with probability 0.75, from a Gaussian distribution with a mean of 5 integer years and a standard deviation of 1, i.e., the initial observations remain at baseline with probability 0.25.  Additionally, if a patient transitions to death prior to the generated initial observation time, then the patient is not included in this {\it delayed enrollment} study.  This is a critical point because it is the cause of the bias: the {\it delayed enrollment} study is less likely to include patients with immediate adverse reactions to heart transplant.  If the sample were truly representative, then all 2000 patients would be represented.  However, studies typically only sample from the living.  The average sample size of the 100 synthetic data sets for the {\it delayed enrollment} study is 1686.

\begin{figure}[t]
\centering
  \vspace{-.15in}
\includegraphics[trim=0cm 9cm 0cm 0cm, clip=true, scale=0.55,page=1]{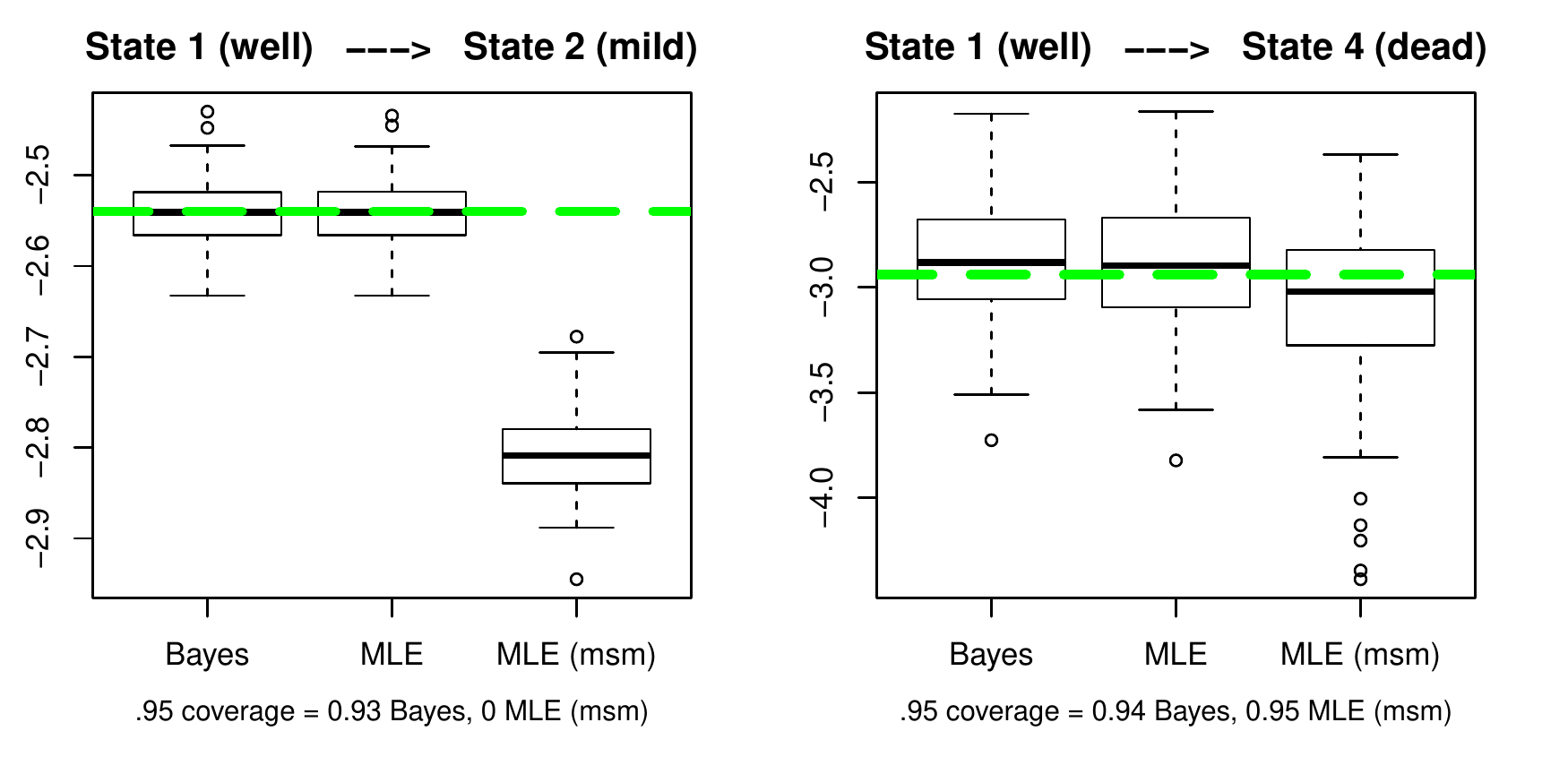}\includegraphics[trim=0cm 9cm 9cm 0cm, clip=true, scale=0.55,page=2]{qmat_intercept_pop}
\includegraphics[trim=0cm 0cm 0cm 9cm, clip=true, scale=0.55,page=1]{qmat_intercept_pop}
\vspace{-.2in}
\caption{Intercept coefficient estimates for a {\it delayed enrollment} study. `Bayes' corresponds to the Bayesian posterior mean estimates, `MLE' corresponds to the MLE obtained via optimizing the likelihood function from Section~\ref{LikelihoodFun} using the `optim' function in $R$, and `MLE (msm)' corresponds to maximum likelihood estimates computed from the `msm' function.  Green dashed lines represent the true values.  Coverage is the proportion of .95 probability credible intervals (confidence intervals for the MLE) which contain the true parameter value.}\label{cavSim_population}
\end{figure}

In Section \ref{delayed_enrollment_section} the procedure for accounting for the {\it delayed enrollment bias} in the likelihood function was described.  It amounts to evaluating the transition rate matrix, $\bQ$, (here, annually) from baseline to initial observation, and then computing the conditional probabilities for the initial states of enrollment.  This conditioning feature is not available in the \verb1msm1 package which was not designed for a {\it delayed enrollment} study.  Figure \ref{cavSim_population} illustrates the effect of ignoring this feature and estimating as though enrollment is $not$ conditional on being alive, i.e., MLE (msm).  The estimates from Bayes and MLE are analogous to those in Figure~\ref{cavSim}, however, they do explicitly account for the initial probability according to (\ref{eq:initial_prob}).  Without accounting for the {\it delayed enrollment} effect in the likelihood function certain estimates are significantly biased downward, suggesting slower rates of transition.  This is because, of the 2000 patients, those which happened to transition quickly through the state space are no longer observed in the data set.  The biases become more extreme as fewer patients are observed at baseline; recall that about 25\% of the patients are still observed at baseline in this example.

The bias also filters into other HMM parameter estimates; see the Supplementary Material for the full results of these two simulation setups.  The objective of this synthetic {\it delayed enrollment} study example is to demonstrate one of the crucial reasons why the analysis of the MCSA data requires methodological developments which are not readily available in standard software.  The \verb1msm1 package, as well as other similar software, are not flawed, rather they were simply not designed for this type of application.

\vspace{-.25in}
\section{Synthetic Mayo Clinic Study of Aging data}\label{demData}
\vspace{-.09in}

Section~\ref{cavData} presented the results of the proposed estimation procedure in a simple idealized simulation example.  In this section a more realistic simulation is presented which is intended to closely replicate the MCSA data generating process with respect to sample size, the frequency of clinical visits, and the proportion of biomarker measurements available.  The three objectives are to (i) provide evidence that the synthetic data reasonably resembles the real data in an effort to verify that the data generating mechanisms of the real data are sufficiently understood, (ii) validate the estimation procedure by demonstrating that credible regions concentrate around the true parameter values, and (iii) demonstrate the reliability of the estimates with respect to the true parameter values.

The standard $R$ package for estimating an HMM with arbitrary continuous-time observations is \verb1msm1 (\cite{Jackson2011}).  While \verb1msm1 is a well written and powerful package, it does not offer Bayesian estimation options, and as discussed previously it cannot correct for the biases on the transition rates that arise from a {\it delayed enrollment} sample. 

The same techniques used to generate data resembling the CAV data set are applied here, just with more features to simulate such as the additional response functions and a {\it death rate bias}.  See the Supplementary Material for the particulars of how the synthetic data was generated.  Before diving right into the estimation results, a few estimation details are discussed, mainly relating to prior information.  

In the state space of Figure \ref{StateSpace}, there are a number of assumptions which can reasonably be made about the biology of this process.  For instance, the rate of transition to the N+ state with high amyloid burden should be at least as fast as with low amyloid burden.  That is, the rate parameter from $\bQ$ for transitioning from A+N$-$ to A+N+ should be at least as big as the rate parameter for transitioning from A$-$N$-$ to A$-$N+.  Similar constraints should be true for transitioning from low to high amyloid burden with respect to high/low neuro-degeneration burden (i.e., having N+ cannot lower the rate of transitioning to A+), and those with dementia die at a rate no less than those without dementia.  These constraints are a mathematical formulation of the assumption that having a larger burden may not escalate disease progression, but it certainly cannot help.  A table with the full list of rate constraints is given in the Supplementary Material.

In addition to these constraints, it is reasonable to assume that all of the age coefficients on the rate parameters (see equation (\ref{logRates})) are nonnegative.  That is, becoming older will not slow down an individual's rates of progression.  Next, constraints are placed on the response variable parameters to help with identifiability.  The mean $\log(\text{PIB} - 1)$ measurement associated with the low amyloid burden state should not exceed that of the high amyloid burden state.  Analogously, the mean Thickness measurement associated with high neuro-degeneration burden should be less than or equal to that for low neuro-degeneration burden (low Thickness is associated with high burden).  Lastly, the mean MMSE scores should be monotone non-increasing in the state ordering $\{4, 5, 6\}$, the mean MMSE score for state 1 should be no smaller than that for states 2 and 3, and the mean MMSE score for states 2 and 3 should be no smaller than that for state 4.  No constraint is placed between the mean scores for states 2 and 3.

One of the advantages of the Bayesian approach of estimation is that imposing this long list of important constraints in the model can be easily handled through specification of the priors, and the priors have a natural ability to accommodate other known information about the model parameters.  In particular, the MCSA sampled subjects are from the greater Rochester, MN area, and Minnesota death rates for women and men of all ages (from 1970 to 2004) are made available in the US Decennial Census, which are captured in the ``survexp.mn'' data set in the \verb1survival1 package in $R$ \citep{Therneau2015}.  These gender-specific rates are used to set the prior means placed on the baseline and gender coefficients for the log death transition rate (equation (\ref{logRates_death})), see Table \ref{priors}.  It is assumed that all non-dementia to death rates are the same (and hence share the same coefficients in equation (\ref{logRates_death})).  The fact that the A and N states do not have obvious external manifestations makes this assumption reasonable.    

A multivariate normal prior is placed on the vector of HMM parameters, and multivariate normal proposals are used.  The full table of prior distribution specifications for each of the 81 parameters in the model is provided in Table \ref{priors}.  For parameters which are non-negative, such as the variance parameters for the response functions, the Gaussian priors are placed (and Gaussian proposals are made) on the natural logarithm of the parameter.  For parameters which are constrained to be between 0 and 1, such as the dementia misclassification parameters and the initial state probability parameters, the logit transformation of the parameters is used.  For example, the initial state probability vector is re-expressed as
\vspace{-.15in}\begin{equation}\label{logit_pi}
\b\pi_{0} = 
\left[1 \:,\: e^{\xi_{1}} \:,\: e^{\xi_{2}} \:,\: e^{\xi_{3}} \:,\: 0 \:,\: 0 \:,\: 0\right]'
\frac{1}{1+e^{\xi_{1}}+e^{\xi_{2}}+e^{\xi_{3}}}
\vspace{-.15in}\end{equation}
where $\xi_{1}$, $\xi_{2}$, and $\xi_{3}$ are assigned Gaussian priors and proposals.  

Although commonly done in the literature (mostly due to conjugacy), we hesitate to use inverse-gamma priors on variances for reasons discussed in \cite{Gelman2006}; namely, the inverse-gamma($\varepsilon$,$\varepsilon$) family of priors is very sensitive to the choice of $\varepsilon$, which does not naturally lend itself as weakly-informative nor uninformative.  Moreover, we view our priors in general as weakly-informative in the sense of \cite{Gelman2006}.  That is, they are set intentionally weaker than what we believe the expert domain knowledge warrants, with the exception of the priors for parameters associated with the non-dementia to death rate and the {\it death rate bias}.  Further, without having more knowledge about the shape of the prior distributions, the symmetry, exponential tails, and mathematical simplicity of Gaussian priors make them a natural choice.  They allow us to be as diffuse as we believe appropriate without affording a questionably large amount of mass at the extreme values of the distributions.  We investigate/discuss sensitivities of our posterior estimates to specifications of our priors in Section \ref{prior_sensitivity}.

\begin{table}[t]
\footnotesize
\renewcommand{\arraystretch}{1.1}

\begin{tabular}{l l l l l l l l}
\multicolumn{8}{c}{{\bf Prior means and standard deviations}} \\
\vspace{-.05in} \\
\multicolumn{8}{l}{\bf State transition parameters (see (\ref{logRates}) and (\ref{logRates_death}))} \\
Transition & $\beta_{0}^{(l)}$ & $\beta_{1}^{(l)}$ & $\beta_{2}^{(l)}$ & $\beta_{3}^{(l)}$ & $\beta_{4}^{(l)}$ & $c$ & $c + d$ \\
\cline{1-8} 
non-dem$\to$7 & -4.41 (.1) & .094 (.01) & .47 (.05) & 0 (.1) & 0 (1) & -.75 (.375) & -.60 (.3) \\
5$\to$7 and 6$\to$7 & -4 (1) & .1 (.05) & 0 (1) & 0 (.1) & 0 (1) & & \\
all others & -3 (1) & .1 (.05) & 0 (1) & 0 (.1) & 0 (1) & & \\
\end{tabular}

\begin{tabular}{l l l l l l l l}
\vspace{-.05in} \\
\multicolumn{8}{l}{\bf Cubic spline parameters for state 1 to state 2 as a function of age} \\
$c_{1}$ & $c_{2}$ & $c_{3}$ & $c_{4}$ & $c_{5}$ & $c_{6}$ & $c_{7}$ & $c_{8}$ \\
\cline{1-8} 
-5 (1) & -4 (2) & -3 (2) & -2 (2) & -1 (2) & 0 (3) & 1 (3) & 2 (3) \\
\end{tabular}

\begin{tabular}{l l l l l l l l l}
\vspace{-.05in} \\
\multicolumn{3}{l}{\bf $\log(\text{PIB}-1)$ response (see (\ref{pibEq}))} & & & \multicolumn{3}{l}{\bf Thickness response (see (\ref{thickEq}))}\\
$\mu_{A-}$ & $\mu_{A+}$ & $\log(\sigma_{\pib}^{2})$ & & & $\mu_{N-}$ & $\mu_{N+}$ & $\log(\sigma_{\thick}^{2})$   \\
\cline{1-3} \cline{6-8}
-1.3 (.2) & -.5 (.2) & $\log\big((.4/3)^{2}\big)$ (2) & & & 3.14 (.2) & 2.34 (.2) & $\log\big((.4/3)^{2}\big)$ (2) \\
\end{tabular}

\begin{tabular}{l l l l l l l l}
\vspace{-.05in} \\
\multicolumn{8}{l}{\bf MMSE response (see (\ref{mmseEq}))} \\
$\alpha_{1}$-$\alpha_{4}$ & $\alpha_{5}$-$\alpha_{6}$ & $\alpha_{7}$ & $\alpha_{8}$ & $\alpha_{9}$ & $\alpha_{10}$ & $\alpha_{11}$ & $\log(\sigma_{\mmse}^{2})$ \\
\cline{1-8} 
-.28 (.75) & -7.3 (3) & 0 (1) & 0 (1) & 0 (1) & 0 (1) & 0 (1) & -.7 (2)  \\
\end{tabular}

\begin{tabular}{l l l l l l l l}
\vspace{-.05in} \\
\multicolumn{3}{l}{\bf Dem misclass (see (\ref{demEq}))} & & & \multicolumn{3}{l}{\bf Initial probabilities (see (\ref{logit_pi}))} \\
$\text{logit}(p_{0})$ & $\text{logit}(p_{1})$ & & & & $\xi_{1}$ & $\xi_{2}$ & $\xi_{3}$ \\
\cline{1-3} \cline{6-8} 
-3 (1) & -3 (1) & & & & -3.5 (.25) & -6 (1) & -6 (1) \\
\end{tabular}\caption{Displayed are the means and standard deviations of the (independent) normal priors placed on the 81 model parameters.  Standard deviations are in parentheses, and these priors assume that the data have been centered.  The parameters $c_{1}, \dots, c_{8}$ are the control points used to estimate the cubic spline for the state 1 to state 2 transition (for baseline and age), as discussed at the end of Section \ref{ContTimeTrans}.}\label{priors}
\end{table}

The simulation study consisted of simulating 50 synthetic data sets resembling the MCSA data.  The synthetic data sets were simulated to contain similar amounts of information to the real data set.  That is, 4742 subjects were simulated starting from random ages and assigned other covariates randomly from the empirical distribution of the MCSA data.  The simulated subjects are ``observed'' at times which are determined by sampling from the actual inter-observations times in the MCSA data.

The maximum length of time in the study for subjects in the MCSA is not much longer than 12 years.  Thus, for reasonable comparison, subjects in the synthetic data sets are observed for 12 years or until time of death, whichever comes first.  Exactly 2718 (approximately 57\%) of subjects in the MCSA data set have at least one observed biomarker measurement, and both biomarkers were observed for 1740 subjects.  Of the 2718 subjects, the proportions of observed biomarkers over all visits is presented in the following table,
\begin{center}
\vspace{-.2in}
\footnotesize
\begin{tabular}{c c | c c }
\multicolumn{2}{r}{} & \multicolumn{2}{c}{PIB} \\
& & measured & not measured \\
\cline{2-4} 
\multirow{2}{*}{Thickness} & measured & 0.227 & 0.231 \\
& not measured & 0.002 & 0.540 \\
\end{tabular}$\;$.
\end{center}
To keep consistent with this feature, $\sim$43\% of the synthetically generated subjects were given no biomarker data for any of their visits, while the remaining $\sim$57\% were randomly given observed PIB or Thickness measurements, at each clinical visit, according to the above distribution.

Death was observed for just over 28\% of the actual MCSA study subjects, and the number of clinical visits for study subjects varied between 1 and 10 visits, with a median of 4.  The synthetic data sets observe death for $\sim$31\% of subjects on average, and the number of clinical visits for synthetic study subjects varies between 1 and 11 visits, with a median of around 6.

\begin{figure}[t]
  \vspace{-.15in}
\centering
\includegraphics[scale=0.45,page=1]{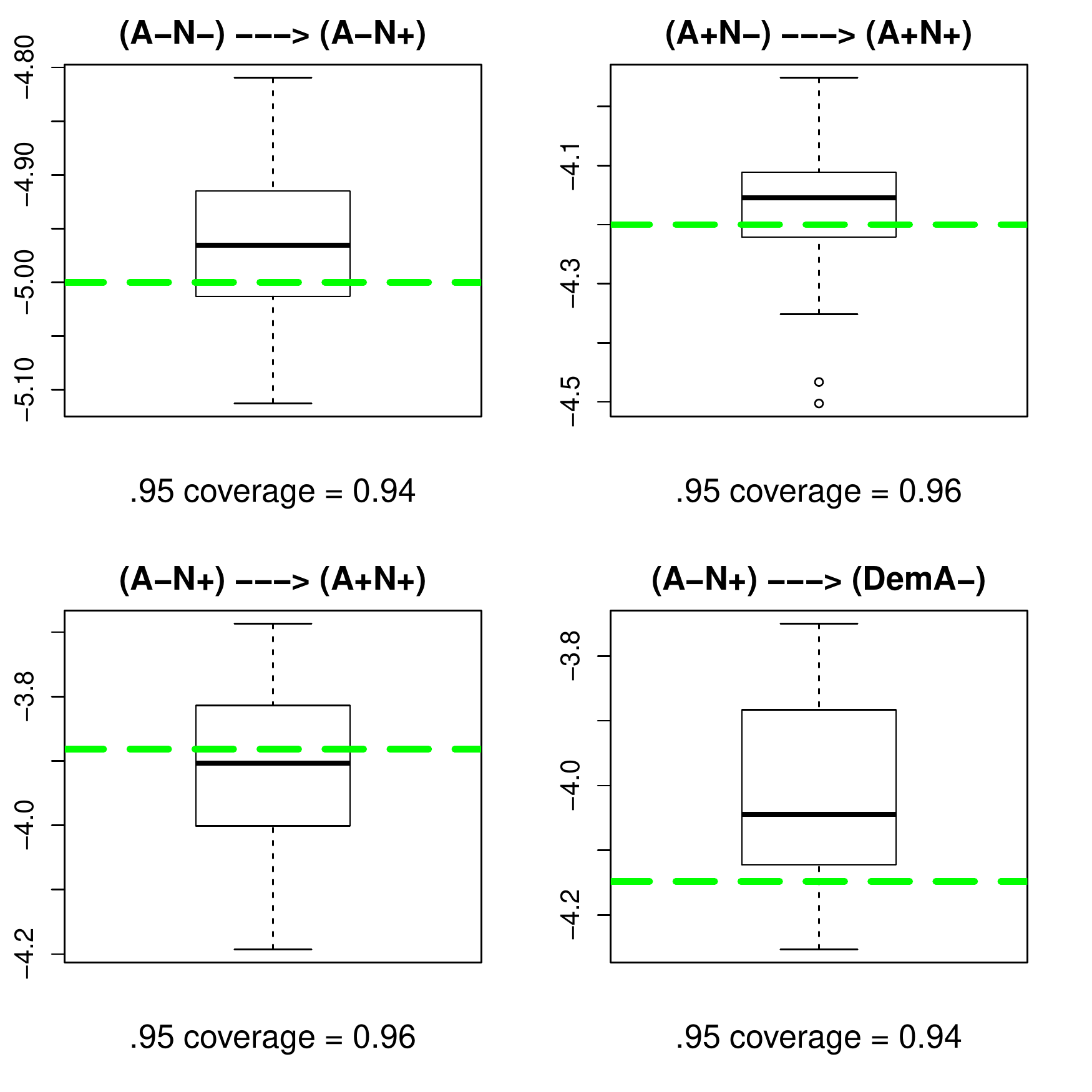}\includegraphics[scale=0.45,page=2]{boxplot_qmat_intercept}
\includegraphics[trim=0cm 9cm 0cm 0cm, clip=true, scale=0.45,page=3]{boxplot_qmat_intercept}\includegraphics[trim=0cm 0cm 0cm 9cm, clip=true, scale=0.45,page=3]{boxplot_qmat_intercept}
\caption{Intercept and {\it death rate bias} coefficient estimates for the synthetic MCSA data, see (\ref{logRates}) and (\ref{logRates_death}).  Note that the covariates in the data are centered.  Presented are box plots of posterior means of the labeled parameters, from 50 synthetic MCSA data sets.  Green dashed lines represent the true values.  Coverage is the proportion of .95 probability credible intervals which contain the true parameter value.
}\label{demSim_age}
\end{figure}

Figure \ref{demSim_age} provides a summary of the results in the form of box plots of the posterior mean estimates, and coverage for 95 percent credible intervals, for 11 of the more interesting model parameters.  The Supplementary Material contains a more detailed summary of the results, including box plots, coverages, histograms, and MCMC trace plots for all 81 model parameters.  Overall, this simulation exercise lends some confidence to the results of the actual MCSA analysis presented next.

\vspace{-.25in}
\section{Analysis of the Mayo Clinic Study of Aging Data}\label{RealDataMayo}
\vspace{-.09in}

This section presents analysis and interpretation of the HMM parameter estimates of the actual MCSA data.  Since there are 81 HMM parameters, some playing very different roles, presenting the output in a concise manner is challenging and depends on the question being asked.  We present a targeted summary of a few important questions here.  See the Supplementary Material for estimates of all 81 HMM parameters, including the MCMC trace plots, histograms, and 95 percent credible intervals.  We parallelize our likelihood computation (within each MCMC step) over 30 threads on a computing cluster, and for 10 unique random number generator seeds we run the MCMC algorithm (i.e., using a total of 300 threads) for about 3 days on the real MCSA data set.

\begin{figure}[b!]
  \vspace{-.0in}
\centering
\includegraphics[scale=0.65]{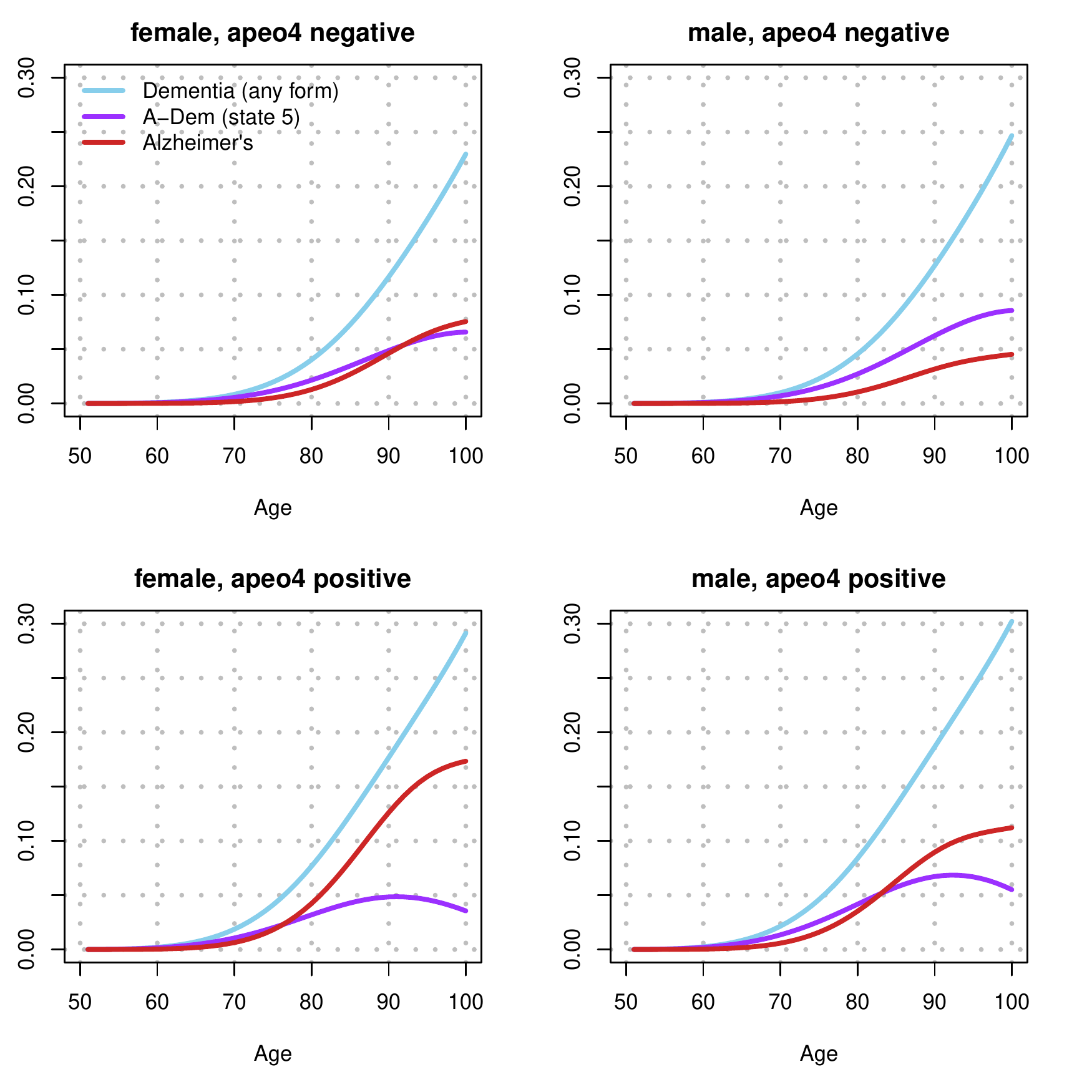}
\vspace{-.2in}
\caption{Evolution of transition probabilities.  The curves represent the probability of transitioning to the respective state for each given age, computed using the posterior mean estimates of the HMM parameters.  The probabilities are conditional on not transitioning to state 7 (Dead), and correspond to an individual in state 1 (A$-$N$-$) at the baseline age of 50.  The label `apoe4 negative' corresponds to an individual with no APOE-$\varepsilon$4 alleles, and `apoe4 positive' corresponds to an individual with at least one APOE-$\varepsilon$4 allele.  The curve labeled `Alzheimer's' depicts the probability of making the transition from state 4 (A+N+) to state 6 (A+Dem), given not dead.}\label{probEvol}
  \vspace{-.2in}
\end{figure}

The state space as it is defined in Figure \ref{StateSpace} allows for the computation of transition probabilities broken down for particular types of dementia.  Most notably, the development of Alzheimer's Disease as defined here corresponds to a transition from state 4 (A+N+) to state 6 (A+Dem), i.e., there was Amyloid build up, prior to the neuro-degeneration leading to Dementia.  See Figure \ref{probEvol} for the estimates of how these transition probabilities evolve over time.  While Alzheimer's Disease is slightly more prominent among females of a given age versus males of the same age, it is interesting that the likelihood of dementia (of any kind) is nearly the same, i.e., males are more likely to develop non-Alzheimer's related dementia.

\begin{wrapfigure}{r}{.49\textwidth}
\centering
\vspace{-.13in}
\includegraphics[scale=0.44]{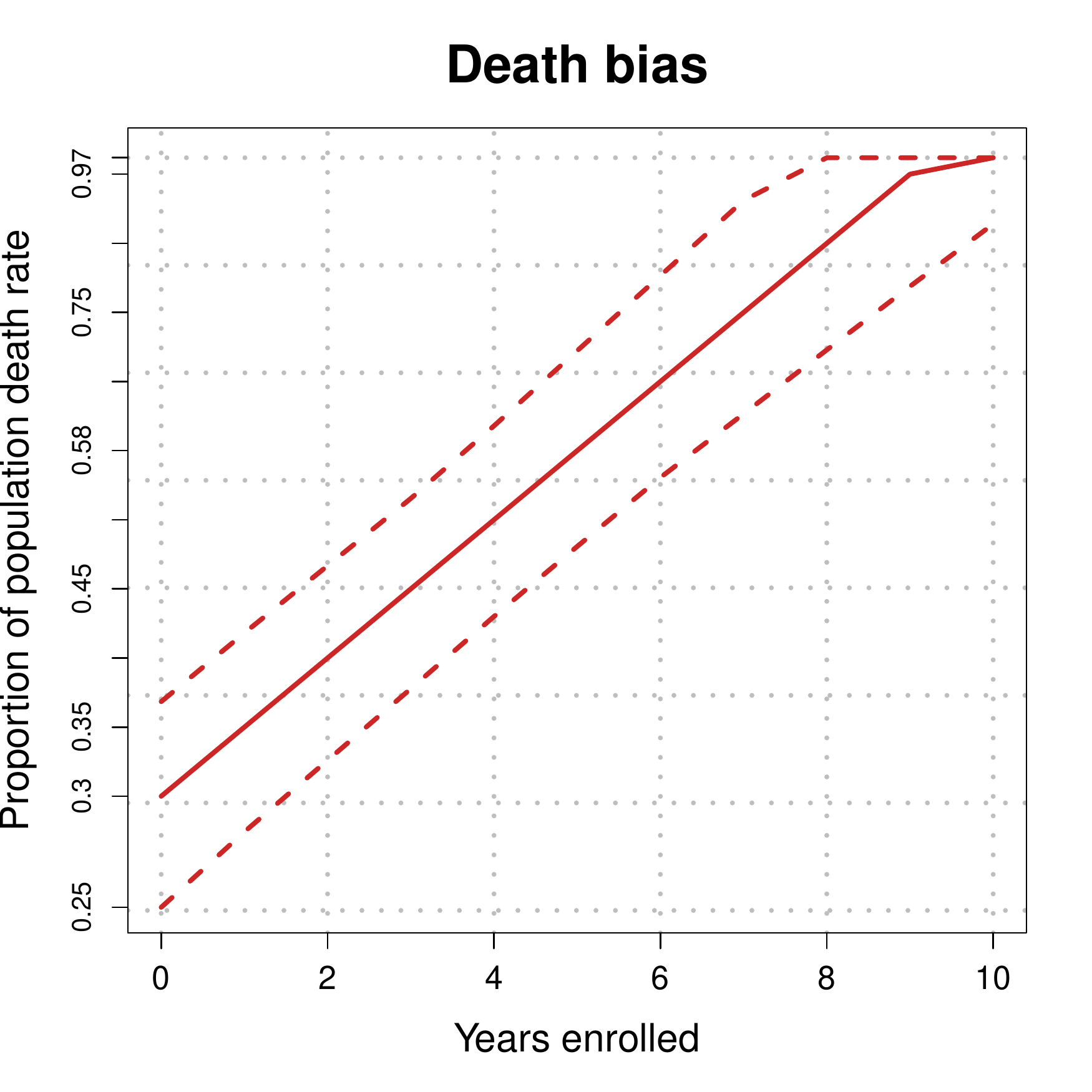}
\vspace{-.4in}
\caption{Posterior mean estimate of the {\it death rate bias} (\ref{logRates_death}).  Values are interpreted as the proportion of the population death rate that is experienced by subjects in the MCSA, for each integer year a subject is enrolled in the study.  For example, subjects just enrolled in the study experience a death rate which is 31 percent of the population death rate.  The dotted lines are 95\% credible bands.}\label{deathBiasSpline}
\vspace{-.2in}
\end{wrapfigure}

The estimated {\it death rate bias} for this study is also of interest, particularly due to the novel approach taken to account for it.  It is observed that individuals just enrolled in the study experience a death rate which is $\sim$31 percent (posterior mean) of the population death rate and it remains lower than the rest of the population for several years after enrollment; See Figure \ref{deathBiasSpline}.  This suggests that the {\it death rate bias} cannot be ignored.   

Another feature of the analysis is that it is cut-point agnostic, by design.  Instead of hardwiring cut-points, the suggested cut-points in the medical literature have been used as prior information for the response distributions from high/low amyloid burden and high/low cortical thickness loss burden in the $\log(\text{PIB} - 1)$ and Thickness measurements, respectively.  Recall, Figure \ref{pibThicknessDensity} displayed the estimated distribution of these response biomarkers for high/low burden states.

\begin{table}[t!]
\footnotesize
  \renewcommand{\arraystretch}{1.1}
  \centering
\begin{tabular}{c c | c c }
\multicolumn{2}{r}{} & \multicolumn{2}{c}{Observed status} \\
& & Diagnosed not demented & Diagnosed demented \\
\cline{2-4} 
\multirow{2}{*}{True status} & Not demented & 0.992   & 0.008 [0.007, 0.009] \\
& Demented & 0.107 [0.072, 0.152] & 0.893  \\
\end{tabular}
\vspace{-.07in}
\caption{\footnotesize Posterior mean estimates of the dementia diagnosis response parameters.  The components are probabilities. Note that each row corresponds to only one parameter, but both columns have been filled in for ease of interpretability (rows must sum to one).  The brackets represent 95 percent credible intervals (for the components directly corresponding to the parameters which were estimated).}\label{demMisclass}
\renewcommand{\arraystretch}{.65}
\vspace{-.1in}
\end{table}

Table \ref{demMisclass} shows the estimated dementia misclassification probabilities.  These estimates indicate that physicians tend to be conservative in diagnosing dementia.  That is, they very seldom diagnose an individual without dementia as demented, but about 1 out of 10 individuals that truly have dementia is not diagnosed as such.

\begin{figure}[t!]
  \begin{center}
\vspace{-.175in}
\includegraphics[scale=.55,page=11]{heatMaps_no_scale}\\[-.3in]
\includegraphics[trim=0cm 7cm 0cm 8cm, clip=true, scale=.4]{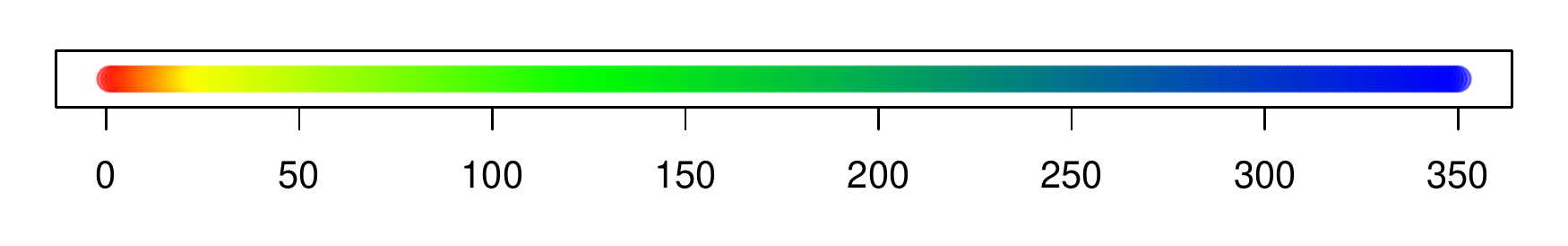}
\vspace{-.2in}
\caption{\footnotesize Posterior mean estimates of the $reciprocal$ transition rate components of $\bQ$ at age 60.  The numerical values can be interpreted as the estimated mean times (in years) to transition, conditional on age.  These plots correspond to an individual with a college degree.  Recall that transitions to dead from states 1-4 are constrained to be equal, and so for ease of presentation only one transition arrow is shown.}\label{heatMap1}
  \end{center}
\end{figure}
\begin{figure}[h!]
  \begin{center}
\includegraphics[scale=.55,page=31]{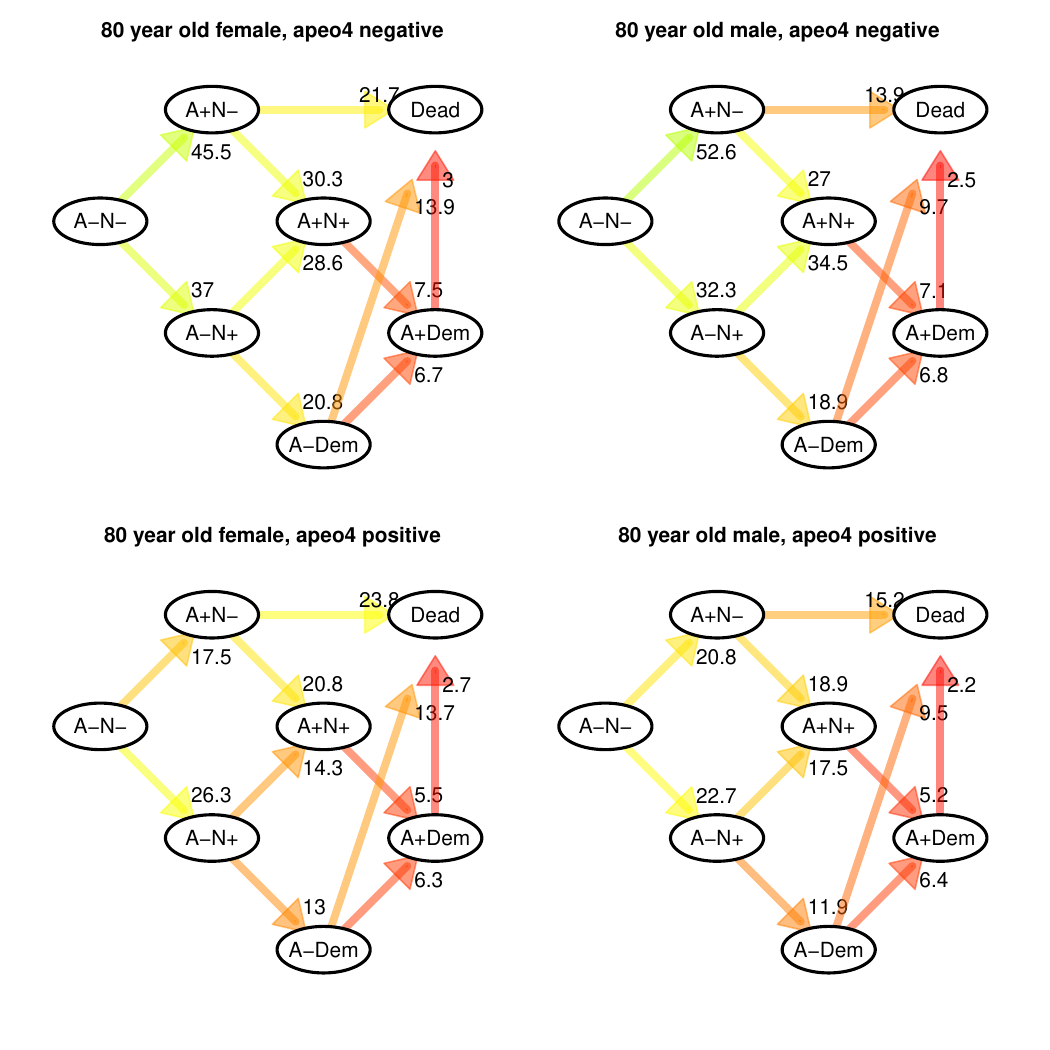}\\[-.3in]
\includegraphics[trim=0cm 7cm 0cm 8cm, clip=true, scale=.4]{heatLegend_no_scale}
\vspace{-.2in}
\caption{\footnotesize Posterior mean estimates of the $reciprocal$ transition rate components of $\bQ$ at age 80.}\label{heatMap2}
  \end{center}
  \vspace{-.45in}
\end{figure}

Finally, Figures \ref{heatMap1} and \ref{heatMap2} provide heat maps of the state space corresponding to individual specific estimated transition intensities (posterior mean estimates).  These depictions of the state space in relation to the estimated parameters captures a holistic picture of the model as a physical system.  With these plots, it is easy to identify which transition rates are most intense at different ages, and to get a general sense of when rates begin to `heat up'.  In fact, these plots can be created for every integer age greater than or equal to 50, and they are presented as a movie in the Supplementary Material.

There are many variables whose effect is known medically that could be used to validate some of the results.  For example, it is observed that the rates are relatively dormant for the first two decades starting from 50 years old.  After that, transition rates among the first four states seem to increase slower than the transition rates to more advanced states.  Transitions particularly to A+Dem from A+N+ or from A$-$Dem are the most intense over almost all ages.  APOE-$\varepsilon$4 is known to increase risk of A+ by a factor of 2-3.  Based on this, the model should yield a higher rate estimate of A$-$N$-$ (state 1) to A+N$-$ (state 2) and A$-$N+ (state 3) to A+N+ (state 4) among APOE-$\varepsilon$4 carriers than non-carriers.  Examination of Figures~\ref{heatMap1}~and~\ref{heatMap2} illustrates that at both ages and for both men and women, the relationship between both of these transition rates and APOE-$\varepsilon$4 in the model output is exactly as would be expected.  Additionally, the rate of A+N+ to A+Dem should be greater in APOE-$\varepsilon$4 carriers than non-carriers.  This is the case in the model for both ages shown and for both men and women as would be predicted from the known biology.

\vspace{-.15in}
\subsection{Sensitivity to prior densities}\label{prior_sensitivity}
\vspace{-.07in}

As with any Bayesian analysis one must consider the sensitivity of the resulting posterior estimates with respect to the prior specifications.  Since the non-dementia to death rate is known to not deviate far from the overall Minnesota death rate, tight priors serve to untangle ambiguity between the death rate and other parameters in a broad sense, or ``on average''.  It should be noted that the non-demented to death rate and {\it death rate bias} parameters are quite sensitive to the variances placed on their respective priors.  We remark once more that the methodology we propose in Section \ref{death_bias_section} for estimating/correcting the {\it death rate bias} relies on strong prior information available for the overall population death rate.  In investigating the sensitivity of the HMM parameter estimates to the priors for the non-dementia to death rate parameters we find that the only unstable parameter estimates are those associated with the rate itself and the {\it death rate bias}, due to a lack of identifiability.

To study sensitivities to the specification of all priors other than those related to the non-dementia to death rate, we increased all other prior standard deviations by a factor of 10 and re-computed the HMM parameter estimates.  See the Supplementary Material to compare the posterior estimates between the more and less diffuse prior specifications.  It was observed that the estimated posterior distributions of each parameter were largely unchanged, with the exception of the log-cubic spline parameters for the transition from state 1 to state 2, and the initial state probabilities.  This can most likely be attributed to the large amount of flexibility afforded by a cubic spline, and a degree of un-identifiability between the cubic spline and initial state parameters because all other parameter estimates are unaffected.  Moreover, the original tighter priors set in Table \ref{priors} for the initial state probabilities are well-informed and reasonable, according to population data.

\vspace{-.25in}
\section{Conclusions \& Future work}\label{Conclusions}
\vspace{-.09in}

A continuous-time HMM was developed for the analysis of the MCSA data.  Much care was taken to make this model as realistic of an approximation to the actual data generating process as possible, including the treatment of important features such as {\it delayed enrollment} and {\it death rate bias}.  A Bayesian computational framework was developed in order to facilitate computation and quantification of uncertainty, as well as allow for essential prior information on many of the parameters.  The model and its estimation performance was validated via several simulation studies, prior to presenting the results of its application to the MCSA data.  Several important findings were that (i) the {\it delayed enrollment} and {\it death rate bias} play a significant role in this study, (ii) females of a certain age are more prone to Alzheimer's related dementia than male counterparts, but they are less prone to dementia, in general, and (iii) individuals with at least one APOE-$\varepsilon$4 allele are more than twice as likely to develop Alzheimer's than those with no alleles.

Our work builds on the simpler Markov model of the MCSA from \cite{Jack2016}, and could be viewed as a competing model.  We include the additional covariates for gender, education, and presence of an APOE-$\varepsilon$4 allele on the state transition rates, and introduce the emitted response variables associated with amyloid and cortical thickness to allow for agnostic biomarker cut-points.  Additionally, we consider the emitted response variables for MMSE scores and physician diagnosis misclassification of dementia, as well as provide methodology for estimating/correcting the {\it death rate bias}.  The model from \cite{Jack2016} used fixed death rates from Minnesota population data.  These added features of our model allow for greater flexibility in the theorized data generating model, and deeper understanding of the biology.  Admittedly, though, if the proposed model became much more complex with additional covariates (or other features) it would require some decisions about which covariates to include in which equations.  Stochastic search variable selection is a viable option as it has been used successfully on event rate models \citep{Storlie12a, George1993}.  Other options include spike and slab priors \citep{Ishwaran2005}, or computing competing models and comparing them with some criterion such as the Widely Applicable Information Criterion (WAIC, see \cite{Watanabe2010}). 

One limitation of the proposed approach is that it does not allow for rates to vary depending on the extent of Amyloid (or cortical thickness) burden.  It is known that once there is sufficient Amyloid build up, then the rate of neuro-degeneration is elevated, however, this effect may not be constant across all levels of Amyloid build up.  This could be tested by allowing for multiple discrete Amyloid states (e.g., low, medium, high).  However, this feature would be best addressed with a continuous-state space for both Amyloid and cortical thickness burden.  This is a subject of future work.

An additional remark is that the various constraints placed within our model such as on the transition rates and on the response function parameters are rather stringent, but they reflect expert domain knowledge.  Moreover, constraints such as those on the PIB and Thickness mean parameters simply make the parameters identifiable.  Nonetheless, imposition of such constraints should not be decided ad hoc.  And when concerns arise as to the reasonability of various constraints, sensitivity tests should be performed to more fully understand the implications.  A similar remark applies to prior specifications, and as we noted in Section \ref{prior_sensitivity} we find that our estimated model, particularly for parameters associated with the death rate and {\it death rate bias}, are heavily reliant on the availability of strong prior information for population death rates.

Lastly, it is likely the case that our computations could be made more efficient by implementing Hamiltonian Monte Carlo methods.  This paper lends credibility to and provides verification for our theorized model and features of the MCSA data.  Accordingly, a natural next step is to sharpen our computational strategies which would facilitate deeper explorations of the features of the data.

{\small
\singlespacing
\bibliographystyle{agsm}
\bibliography{References}
}

\end{document}